\documentclass[aps,prc,twocolumn,groupedaddress,showpacs]{revtex4}

\usepackage{graphicx}
\usepackage{color}
\usepackage{epsfig}
\usepackage{amsmath}
\newcommand{\2}{^{II}}
\begin{document}

\title{Conditions for Phase Equilibrium in Supernovae, Proto-Neutron and Neutron Stars}

\author{M. Hempel, G. Pagliara, and J. Schaffner-Bielich}
\affiliation{Institute for Theoretical Physics, Ruprecht-Karls-University, Philosophenweg 16, 69120 Heidelberg, Germany}

\date{\today}

\begin{abstract}
We investigate the qualitative properties of phase transitions in a general way, if not the single particle numbers of the system but only some particular charges like e.g.~baryon number are conserved. In addition to globally conserved charges we analyze the implications of locally conserved charge fractions, like e.g.~local electric charge neutrality or locally fixed proton or lepton fractions. The conditions for phase equilibrium are derived and it is shown, that the properties of the phase transition do not depend on the locally conserved fractions. Finally, the general formalism is applied to the liquid-gas phase transition and the hadron-quark phase transition for typical astrophysical environments like in supernovae, proto-neutron or a neutron stars. We demonstrate that the Maxwell construction known from cold-deleptonized neutron star matter with two locally charge neutral phases requires modifications and further assumptions concerning the applicability for hot lepton-rich matter. All possible combinations of local and global conservation laws are analyzed, and the physical meaningful cases are identified. Several new kinds of mixed phases are presented, as e.g.~a locally charge neutral mixed phase in proto-neutron stars which will disappear during the cooling and deleptonization of the proto-neutron star.
\end{abstract}

\maketitle
\section{Introduction}
Phase transitions (PTs) of the first kind play a crucial role in various astrophysical systems like cold neutron stars, proto-neutron stars, collisions of compact stars or supernovae. In all of these systems the equation of state (EoS) is one of the key information for the understanding of the various phenomena connected with them. For hydrodynamic and hydrostatic calculations the EoS has to be known, and therefore the correct treatment of phase transitions is essential.

In fact plenty of phase transitions can occur in the aforementioned systems: At densities below saturation density and temperatures lower than $\sim 15$ MeV the well-known liquid-gas phase transition of nuclear matter occurs, leading to a dense symmetric phase in coexistence with a more dilute neutron gas \cite{ravenhall83,lamb83,muller95,bugaev01,ishizuka03}. At several times saturation density further phase transitions to exotic degrees of freedom like to quark matter in the chiral phase transition \cite{itoh70,heiselberg93} or to a kaon condensed phase \cite{fujii96, glendenning98,glendenning99,pons00} are possible. A phase transition to a pion condensed phase \cite{migdal79,haensel82,migdal90} or to hyperon matter \cite{schaffner02} might also be of first order. Phase transitions between different types of color superconducting quark matter were proposed in Refs.~\cite{Ruster06nourl,blaschke06,pagliara08}.

The inclusion of a phase transition to exotic degrees of freedoms can substantially alter the stability of the star. Usually, the phase transition leads to a softening of the EoS and therefore lowers the maximum mass which can be supported by the star.  Thus, the theoretical predictions can be confronted with real observations of neutron star masses. Besides the mass, also other observables can be linked to phase transitions in compact stars, e.g.~gamma ray bursts \cite{fryer98,mishustin03,berezhiani03,drago08} or sudden spin ups during the rotational evolution of young pulsars \cite{glendenning97,zdunik06}. Furthermore the appearance and the structure of mixed phases can have important consequences for transport properties like the thermal conductivity or the neutrino emissivities and opacities \cite{reddy00}. Also the shear modulus and the bulk viscosity will be altered, affecting the glitch phenomena or r-modes \cite{glendenning01,blaschke01}. Consequently, the occurrence of mixed phases can modify the thermal \cite{page2006} and rotational evolution of compact stars.

The description of phase transitions in cold, deleptonized neutron stars is rather well understood and extensively discussed in the literature. For bulk matter there are two, in principle different, possible treatments: the Maxwell construction based on local charge neutrality leading to a first order phase transition or the Gibbs conditions for phase equilibrium where only global charge neutrality is required. For the latter case, which was first discussed in \cite{glendenning92}, the phase transition becomes second order according to Ehrenfest's classification and an extended mixed phase appears inside the neutron star \cite{muller97}. If one wants to go beyond the bulk limit, finite size effects in form of surface and Coulomb energies need to be included, as was done in \cite{heiselberg93,voskresensky03,maruyama07,maruyama08a, maruyama08b} for a mixed phase of (hyperonic) hadronic matter and quark matter. Although Gibbs conditions are used in most of the publications about the hadron-quark phase transition in the bulk approximation, as e.g.~in Ref.~\cite{pons01}, these detailed calculations show that the EoS of the phase transition is in fact more similar to the Maxwellian case, if Coulomb and surface contributions are included. Because of large surface tensions and small Debye screening lengths charge screening effects are large, and the phases become almost charge neutral. Then also global properties, like the mass-radius relation of cold neutron stars, resemble more the results of the Maxwell construction. Already in \cite{heiselberg93} it was estimated, that for $\sigma > 70$ MeV/fm$^2$ the Maxwellian case is recovered, recently validated by \cite{maruyama08a}. In \cite{maruyama06} a first order phase transition to a kaon condensed phase was studied and similar results are found. For the case of nuclear mixed phases (liquid-gas PT) the same charge screening effects are observed, but much less pronounced because of the low charge and baryon densities involved \cite{maruyama05}.

Recently, new works about the QCD phase transition in core-collapse supernovae or in compact star mergers appeared, as e.g.~\cite{nakazato08,sagert09} or \cite{bauswein09}. For the proper description of these more complex dynamical scenarios detailed numerical simulations are necessary. Within certain simplifications and/or model assumptions some interesting results were found. E.g.~in Ref.~\cite{sagert09} a successful explosion in spherical symmetry of a 15 M$_\odot$ progenitor star was obtained due to an early appearance of a mixed phase of quarks and nucleons.

For the early stages after the bounce in a supernova, one can represent the evolution of the proto-neutron star by some typical static configurations, see e.g.~\cite{prakash95,prakash97}. Until now, the conditions for phase equilibrium of matter in supernovae and proto-neutron stars were not discussed in detail in the literature. If only global charge neutrality is considered, the situation is clear and completely described in \cite{glendenning92}. However, in some cases the Gibbs conditions for phase equilibrium cannot be fulfilled at all, or only the simple Maxwell construction is wanted for the sake of simplicity, as e.g.~in Ref. \cite{burgio08}, because then no mixed phase is present. But as will be shown in this article, the Maxwell construction has some subtleties and the usual procedure for cold deleptonized neutron stars cannot be used for proto-neutron stars. The requirement of conservation of lepton and/or proton number in addition to baryon number leads to significant differences in the equilibrium conditions, which was not taken into account in several previous publications, like e.g.~in \cite{yasuhira01,nicotra06,yasutake09}. The simple Maxwell construction (in the sense that the mixed phase vanishes) is only possible if in addition to local charge neutrality some other charges are fixed locally, resulting in particular new conditions for phase equilibrium. Because of the additional conserved charges involved, there is a large variety of different descriptions of the phase transition, all of them representing different physical scenarios. We note that one of the new Maxwell constructions which we will present in this article, was already considered in the work by Lugones and Benvenuto \cite{lugones98}.

The structure of this article is as follows: In section \ref{sec_eq} we derive the equilibrium conditions for a general system of two phases in which not the single particle numbers, but only some global charges are conserved. In addition to globally conserved charges, we also consider the case in which some charge fractions are conserved locally. To achieve a general description which can be applied to all the combinations of global and local conservation laws, we start with a formulation, in which all the charge fractions are fixed locally. Then we derive the new equilibrium conditions if some of these local constraints are lifted. In section \ref{sec_proppt} we present the properties of the phase transition depending on whether one has a single- or a multi-component body (only one or multiple globally conserved charges). While most of this was already discussed in \cite{glendenning92} we want to reformulate it by using consistently our formalism and notations. Quite remarkably, we will show that locally conserved fractions do not influence the qualitative behavior of the mixed phase. In section \ref{sec_app} the general results will be applied to single and mixed phases of hadronic matter and/or quark matter. We differentiate between isothermal and adiabatic phase transitions. All relevant possibilities of local and global conservation of the different conserved charges will be analyzed and the corresponding equilibrium conditions will be presented. We will use the results of the previous sections and will show which assumptions are necessary to achieve the simple Maxwell construction even in the case of multiple conserved charges. Some of the combinations of local and global conservation laws have not been discussed in the literature so far, leading to new conditions for chemical equilibrium and new interesting properties of the mixed phase. In addition we discuss the role of neutrinos: We will show explicitly that also in a mixed phase the neutrinos can be handled as an independent contribution to the EoS, as long as the proton fraction is conserved. At the end we will discuss the role of the additional quantum number of strangeness in strange matter, i.e.~three-flavor quark matter or hyperonic matter. In section \ref{sec_sum} we will summarize the main results and draw the conclusions.

\section{Equilibrium Conditions}
\label{sec_eq}
Consider a thermodynamic system with volume $V$ and temperature $T$ composed of two different phases. The numbers of the $N^I$ ($N^{II}$) different particle species of phase $I$ $(II)$ are denoted by ${N^I_i}$ ($N^{II}_j$). To distinguish the two phases we introduce the index $k=I,II$. The thermodynamic potential of the system is the Helmholtz free energy $F(T,V,N_i^k)$.

In many cases, there exist some conserved charges, but the single particle numbers are not conserved. Then it will be more convenient to use these conserved charges as the independent degrees of freedom for the description of the state of the system, instead of specifying all the single particle numbers. In this section the equilibrium conditions in terms of the chemical potentials of the particles shall be derived.

Let us assume there are $C$ conserved charges $C_\sigma$, $\sigma=0, ..., C-1$ for which a conservation law of the following form exists:
\begin{eqnarray}
C_\sigma(N_i^I,N_j\2)&=&C_\sigma^I(N_i^I)+C_\sigma\2(N_j\2)=\textrm{const.}  \; ,\nonumber \\
C_\sigma^k(N_i^k)&=&\sum_i \alpha^k_{i\sigma} N^k_i \label{eq_cg}
\end{eqnarray}
with $\alpha_{i\sigma}^k$ denoting the amount of charge $C_\sigma$ carried by particle $i$ of phase $k$. The total conserved charge consists of the charge in phase I and in phase II.

In addition to the $C$ global conservation laws, we consider additional local constraints, which depend only on the particles in one of the two phases. To be more specific, we require that some (denoted by the index $\tau$, $\tau \neq 0$) of the local charge fractions of the globally conserved charges $C_\sigma$ have to be equal in the two phases:
\begin{eqnarray}
&Y_\tau^I=Y_\tau\2&  \label{eq_y} \\
&Y_\tau^k(N_i^k)=\frac{C_\tau^k}{C_0^k}=\frac{1}{C_0^k}\sum_i \alpha_{i\tau}^k N_i^k= \textrm{const.}& \; ,
\end{eqnarray}
where $C_0^k$ is the local part of the conserved charge $C_0$ and $C_\tau^k$ the local part of $C_\tau$, which is also one of the globally conserved charges. $C_0^k$ shall be a positive, non-vanishing quantity so that it is suitable to characterize the size of the phases in the mixture. If we introduce the global fraction $Y_\tau$,
\begin{equation}
Y_\tau=C_\tau/C_0=\textrm{const.} \; ,
\end{equation}
we find:
\begin{equation}
Y_\tau^k=Y_\tau=\textrm{const.} \; . \label{eq_yeq}
\end{equation}
Eq.~(\ref{eq_yeq}) is the reason why we can evaluate the local constraint of eq.~(\ref{eq_y}) independently for the two phases.

To achieve a general description applicable for all kind of local constraints we will first use the two local parts $C_0^k$ of $C_0$ and all the local charge fractions $Y_\tau^k$, $\tau=1, ..., C-1$,  as the state variables, i.e.~we treat them as independent degrees of freedom. In the following we want to derive the equilibrium conditions for the internal dependent degrees of freedom ${N_i^k}$ if only these state variables are specified $F(T,V,N_i^k)= F(T,V,N_i^k(T,V,C_0^l,Y_\tau^l)$ and are kept constant. Thus:
\begin{eqnarray}
&C_0^k=\textrm{const.}& \label{eq_loc1} \\
&Y_\tau^k= \textrm{const.}& \; . \label{eq_loc2}
\end{eqnarray}
Later we connect these state variables with the real local constraints, in which only some of the local fractions have to be equal and $C_0$ is conserved only globally.

From the first and second law of thermodynamics we get for the free energy $F$ expressed by the particle numbers $N_i^k$
\begin{equation}
0=dF=\sum_{i,k} \left.\frac{\partial F}{\partial N_i^k}\right|_{N^k_{j\neq i},N^{\bar k}_j}dN_i^k \; , \label{eq_df}
\end{equation}
if the volume and the temperature are kept constant. $\bar k$ denotes the phase different to $k$. As the temperature is one of the state variables of the two phases it is set equal by construction, so that thermal equilibrium between the two phases is assured. Only the total volume $V=V^I+V\2$ is kept constant, but the two subvolumes can vary, leading to pressure equality as the condition for mechanical equilibrium:
\begin{equation}
p^I=p\2 \; . \label{eq_p}
\end{equation}

With
\begin{equation}
\mu_i^k =\left.\frac{\partial }{\partial N_i^k}\right|_{N^k_{j\neq i},N^{\bar k}_j}F(N_j^l)
\end{equation}
eq.~(\ref{eq_df}) becomes
\begin{equation}
\sum_{i,k} \mu_i^k dN_i^k=0 \; .
\end{equation}
The constraints of eqs.~(\ref{eq_loc1}) \& (\ref{eq_loc2}) can be implemented by the means of Lagrange multipliers $\lambda_0^k, \lambda_{Y_\tau}^k$, by adding
\begin{eqnarray}
&\lambda_0^k dC_0^k=\lambda_0^k \sum_i \alpha_{i0}^k dN_i^k=0& \; , \label{eq_lambda1a}
\end{eqnarray}
and
\begin{eqnarray}
&\lambda_{Y_\tau}^k dY_\tau^k=\lambda_{Y_\tau}^k \frac{1}{C_0^k}(dC_\tau^k-Y_\tau^k dC_0^k)=0& \nonumber \\
&\Leftrightarrow \lambda_{Y_\tau}^k \frac{1}{C_0^k}\left(\sum_i \alpha_{i\tau}^k dN_i^k-Y_\tau \sum_i \alpha_{i0}^k dN_i^k\right)=0\; ,& \label{eq_lambda1b}
\end{eqnarray}
to $dF$, where we used eq.~(\ref{eq_yeq}). This leads to:
\begin{equation}
\mu_i^k=\lambda_{0}^k\alpha_{i0}^k+ \sum_\tau \lambda_{Y_\tau}^k \frac{1}{C_0^k}\left(\alpha^k_{i\tau}-Y_\tau\alpha_{i0}^k\right) \; . \label{eq_mui1}
\end{equation}

$dF$ can also be expressed as a function of $(C_0^k, Y_\tau^k)$:
\begin{eqnarray}
0=dF&=&\sum _k \left.\frac{\partial F}{\partial C_0^k}\right|_{C_{0}^{\bar k}Y_\eta^l}dC_0^k+ \nonumber \\
&& \sum_{k, \tau} \left.\frac{\partial F}{\partial Y_\tau^k}\right|_{C_{0}^l,Y_\eta^{\bar k},Y_{\eta \neq \tau}^k}dY_\tau^k \; .
\end{eqnarray}
We introduce the chemical potential of the conserved charges $C_0^k$,
\begin{equation}
\mu_0^k=\left.\frac{\partial}{\partial C_0^k}\right|_{C_{0}^{\bar k}Y_\eta^l}F(C_0^l, Y_\eta^l) \; ,
\end{equation}
and of the conserved fractions $Y_\tau^k$:
\begin{eqnarray}
\mu_{Y_\tau}^k=\left.\frac{\partial }{\partial Y_\tau^k}\right|_{C_{0}^l,Y_\eta^{\bar k},Y_{\eta \neq \tau}^k}F(C_0^l, Y_\eta^l) \; .
\end{eqnarray}
Then it is easy to realize, that the Lagrange multipliers are equal to the chemical potentials of the corresponding charges: $\lambda_0^k=\mu_0^k, \lambda_{Y_\tau}^k=\mu_{Y_\tau}^k$, so that:
\begin{equation}
\mu_i^k=\mu_{0}^k\alpha_{i0}^k+ \sum_\tau \mu_{Y_\tau}^k \frac1{C_0^k}\left(\alpha^k_{i\tau}-Y_\tau\alpha_{i0}^k\right) \; . \label{eq_mui2}
\end{equation}

It is interesting to see, that the chemical potentials of the particles depend now directly on the value of the locally fixed charge fractions and the unknown value of $C_0^k$. $1/C_0^k$ appears because a change in $Y_\tau^k$ implies a change in the particle numbers proportional to $C_0^k$. With
\begin{eqnarray}
\mu_{Y_\tau}^k&=&C_0^k \left.\frac{\partial F}{\partial C_\tau^k}\right|_{C_{0}^l,Y_\eta^{\bar k},Y_{\eta \neq \tau}^k} \nonumber \\
&=& C_0^k \mu_\tau^k \; , \label{eq_muy}
\end{eqnarray}
one can see that $\mu_{Y_\tau}^k$ is proportional to $C_0^k$ and the chemical potential of the charge $C_\tau^k$. In fact the local chemical potentials of the particles can only depend on other local intensive variables, so $C_0^k$ has to drop out of eq.~(\ref{eq_mui2}).

By using this result we get the following expression for the equilibrium conditions for the chemical potentials of the particles:
\begin{equation}
\mu_i^k=\mu_{0}^k \alpha_{i0}^k+ \sum_\tau \mu_{\tau}^k \left(\alpha^k_{i\tau}-Y_\tau\alpha_{i0}^k\right) \; . \label{eq_mui}
\end{equation}
The first two of the three terms  simply state that the chemical potential of particle $(i,k)$ is given by the sum over the amount of conserved charges that the particle carries multiplied by the corresponding chemical potentials. The term proportional to $Y_\tau$ appears only for particles which contribute to $C_0^k$. It is due to the change implied by $dC_0^k$ in the charge $C_\tau^k$ if $Y_\tau^k$ is kept constant. It would not appear if instead of the fractions the  chargeswere used for the description of the state of the system. This shows the importance in the definition of the chemical potentials of which other quantities are kept constant.

It is important to realize that the local chemical potentials  $\mu_{0}^k, \mu_{\tau}^k$ will in general be different in the two phases, leading to different chemical potentials of all particles. Only particles of the same phase which carry the same quantum numbers will have equal chemical potentials.

All together there are $2C+N^I+N\2+1$ unknown variables in eq.~(\ref{eq_mui}): the chemical potentials $\mu_0^k,\mu_\tau^k, \mu_i^k$  and one of the two subvolumes $V^k$. They can be determined from the $N^I+N\2$ chemical equilibrium conditions (\ref{eq_mui}), pressure equilibrium (\ref{eq_p}) and the $2C$ conservation laws (\ref{eq_loc1}) and (\ref{eq_loc2}) for the fixed state $(C_0^k,Y_\tau^k)$. If all the relations $N_i^k=N_i^k(T,V^k,\mu_j^k)$ are known, the system is determined completely.

Eq.~(\ref{eq_mui}) can also be applied for a single phase $k$, by setting $V^{\bar k}=0$ for the other phase, equivalent to $N_i^{\bar k}=0$. Then pressure equilibrium is not required any more, and the whole system of equations can be solved and all thermodynamic quantities can be determined, too. If the number of conserved charges $C$ is equal to the number of particles $N^k$ in this phase, then the conserved charges directly fix all the $N^k$ particle numbers $N_i^k$. If $C<N^k$, $N^k-C$ equilibrium conditions between the chemical potentials of the particles will exist.

Next we want to understand the consequences if indeed only $C_0$ but not the $C_0^k$ are conserved and if besides the global conservation of the other charges $C_\tau$, $\tau=1, ..., C-1$ only for the $L$ fractions $Y_\lambda^k$ local constraints of the form (\ref{eq_y}) exist. I.e.~we want to connect the chosen state variables to some particular set of local conservation laws. We will denote the only globally conserved charge fractions by $Y_\gamma$, $\gamma=1, ..., G-1$, with the number of globally conserved charges $G$. For the index $\lambda$ of the locally conserved charge fractions $Y_\lambda^k$ we then have $\lambda=G, ..., C-1$ so that $C=G+L$.

The global constraint for $C_0$ can be written as:
\begin{eqnarray}
\lambda_0 dC_0=\lambda_0 \sum_{i,k} \alpha_{i0}^k dN_i^k=0 \; . \label{eq_lambda2a}
\end{eqnarray}
All the global charge fractions $Y_\gamma=C_\gamma/C_0$ are also conserved:
\begin{eqnarray}
&\lambda_{Y_\gamma} \frac{1}{C_0}(dC_\gamma-Y_\gamma dC_0)=0& \nonumber \\
&\Leftrightarrow \lambda_{Y_\gamma} \frac{1}{C_0}\left(\sum_{i,k} \alpha_{i\gamma}^k dN_i^k-Y_\gamma \sum_i \alpha_{i0}^k dN_i^k\right)=0\; .& \label{eq_lambda2b}
\end{eqnarray}
We already implemented the new Lagrange multipliers $\lambda_0, \lambda_{Y_\gamma}$.

By comparing eqs.~(\ref{eq_lambda1a}) \& (\ref{eq_lambda1b}) with (\ref{eq_lambda2a}) \& (\ref{eq_lambda2b}), one finds that the local conservation laws lead to the same constraints (\ref{eq_lambda2a}) \& (\ref{eq_lambda2b}) which are added to $dF$ if we set:
\begin{eqnarray}
&\lambda_0^I=\lambda_0\2=\lambda_0&  \\
&\frac{\lambda_{Y_{\gamma}}^I}{C_0^I}=\frac{\lambda_{Y_{\gamma}}\2}{C_0\2}= \frac{\lambda_{Y_{\gamma}}}{C_0}&\; ,
\end{eqnarray}
which is equivalent to:
\begin{eqnarray}
&\mu_0^I=\mu_0\2=:\mu_0& \label{eq_condeq1} \\
&\frac{\mu_{Y_\gamma}^I}{C_0^I}=\frac{\mu_{Y_\gamma}\2}{C_0\2}&  \nonumber \\
\Leftrightarrow &\mu_\gamma^I=\mu_\gamma\2=:\mu_\gamma& \; , \label{eq_condeq2}
\end{eqnarray}
where we used eq.~(\ref{eq_muy}) in the last line. Eqs.~(\ref{eq_condeq1}) \& (\ref{eq_condeq2}) are the new additional equilibrium conditions for the globally conserved charges in terms of the local chemical potentials. To express the equality of these chemical potentials we introduced the variables $\mu_0, \mu_\gamma$, which are the global chemical potentials of the corresponding charges. If a local conservation law is lifted, an additional condition for chemical equilibrium between the two phases appears, and the whole set of equilibrium equations can still be solved. Eqs.~(\ref{eq_condeq1}) \& (\ref{eq_condeq2}) are the expected result that local chemical potentials become equal in the two phases when the corresponding local constraint is lifted. Then the two local fractions can adjust to minimize the free energy. After equilibrium is reached, the two phases can be separated from each other without any changes arising.

If one of the globally conserved charges, denoted by $C_\delta$ in the following, is actually not conserved any more this means that
\begin{equation}
\frac{\partial F}{\partial C_\delta}=0
\end{equation}
to minimize the free energy, leading to
\begin{equation}
\mu_{\delta}^I=\mu_{\delta}\2=0 \label{eq_mug0} \; .
\end{equation}
A non-conserved fraction gives two local constraints for the chemical potentials. The two eqs.~(\ref{eq_mug0}) replace the equilibrium condition (\ref{eq_condeq1}). With this additional information the whole system can be determined, even without knowing the value of $C_\delta$ any more. This is in accordance with our previous conclusion that all thermodynamic quantities can be determined, independently of the number of conserved charges $C$ and the number of particles $N^I$ and $N\2$. We note that with the new information of eq.~(\ref{eq_mug0}) the chemical potentials of the remaining conserved charges can possibly be written in a different simplified form.

This procedure for non-conserved charges can also be applied for a single phase $k$, in which only one chemical potential $\mu_\delta^k$ exists, leading to:
\begin{equation}
\mu_{\delta}^k=0 \label{eq_mug0g} \; .
\end{equation}
All other conclusions are also analog to the mixed phase.

From eqs.~(\ref{eq_mui}) it follows immediately, that two particles of the same phase have equal chemical potentials if they carry the same quantum numbers:
\begin{equation}
\mu_m^k=\mu_n^k\; \textrm{if} \; \alpha^k_{m\nu}=\alpha^k_{n\nu} \; . \label{eq_mumneq1}
\end{equation}
If particles $m$ and $n$ of two different phases carry the same quantum numbers, $\alpha^k_{m\nu}=\alpha^{\bar k}_{n\nu}$, e.g.~if some of the particles in the two different phases are identical, this is no longer true in general. The local chemical potentials of local fractions will be different in the two phases. Consequently, if a particle carries global and local charges, its chemical potential will also be different in the two phases. Only if they do not contribute to the locally conserved charges from eqs.~(\ref{eq_mui}), (\ref{eq_condeq1}) \& (\ref{eq_condeq2}) it follows that the chemical potentials of such particles are equal:
\begin{equation}
\mu_m^k=\mu_n^l \; . \label{eq_mumneq2}
\end{equation}
This means that, since such particles can be exchanged freely between the two phases, in equilibrium always the same amount of energy is needed when the number of particles $N_i^k$ is varied in one of the two phases. If no local constraints are applied, the chemical potentials of all identical particles become equal, which are the well-known Gibbs conditions for phase equilibrium.

For fixed temperature $T$, pressure $p$ and particle numbers $N_i^k$, the correct thermodynamic potential is the Gibbs potential $G$, which is also called the Gibbs free enthalpy. With eq.~(\ref{eq_mui}) \& (\ref{eq_condeq1})  we get the following relation inside the mixed phase:
\begin{eqnarray}
G(T,p,N_i^k(C_0,Y_\tau))&=&\sum_{i,k} \mu_i^k N_i^k\nonumber \\
&=&\mu_0 C_0=\mu_0^IC_0^I+\mu_0\2 C_0\2 \; , \label{eq_summua}
\end{eqnarray}
independently of which local constraints are applied. For a single phase $k$ with $C_0^k=C_0$ and $V^k=V$, eq.~(\ref{eq_mui}) leads to:
\begin{equation}
G(T,p,N_i^k(C_0,Y_\tau))=\sum_{i} \mu_i^k N_i^k=\mu_0^k C_0 \; . \label{eq_summub}
\end{equation}
In this case in principle the index $k$ can also be suppressed because only one single phase exists. These two relations can also be used in other thermodynamic potentials or in the fundamental relation of thermodynamics.

\section{Properties of the phase transition}
\label{sec_proppt}
Depending on the number of globally conserved charges the properties of a phase transition are qualitatively different. The differences between a single and a multi component body were extensively discussed in \cite{glendenning92} for neutron stars. In the following we will show that locally fixed charge fractions do not influence the qualitative behavior of the phase transition.

For numerical simulations of proto-neutron stars and supernovae, usually an equation of state in tabular form, as e.g.~\cite{lattimer91nourl,shen98, shen98_2}, in terms of $(T,c_0,Y_{\tau})$, $\tau=1, ..., C-1$, with the density $c_0=C_0/V$ is applied (in these EoSs the state variables are the temperature, the baryon density and the proton fraction $Y_p$). If the volume $V$ (which is completely arbitrary in the thermodynamic limit) is also taken as one of the state variables, all the conserved charges $C_\sigma$, $\sigma=0, ..., C-1$, are known, too, and the Helmholtz free energy becomes the appropriate potential: $F=F(T,V,C_\sigma)$. In the following we want to discuss the properties of the mixed phase for such an EoS in terms of $(T,V, C_\sigma)$ if only $V$ is varied and the other state variables $C_\sigma$ and $T$ are kept constant, i.e.~we are investigating an isothermal change of $c_0$ at constant $Y_\tau$. As before we assume that only $L\leq C-1$ of the fractions are fixed locally in the form: $Y_{\lambda}^I=Y_{\lambda}\2=Y_{\lambda}$. For the globally conserved fractions we will continue to use the index $\gamma$ instead. The number of globally conserved charges/fractions is then $G=C-L$.

By analyzing further the condition for pressure equilibrium one can make statements about the qualitative properties of the phase transition and the possible appearance of mixed phases. The pressure in each phase can only depend on the local chemical potentials of the particles in this phase:
\begin{equation}
p^I(T,\mu^I_i)=p^{II}(T,\mu^{II}_j) \; .
\end{equation}
We already argued before, that also for a single phase the knowledge of the conserved charges, the temperature and the volume $V$ is sufficient to determine all thermodynamic quantities. Obviously, the chemical potentials of the particles cannot depend on the size of the phase, so that they have to be determinable by the local density $c_0^k$ and the local fractions $Y_\tau^k$ alone:
\begin{equation}
p^k(T,\mu_i^k)=p^k(T,c_0^k,Y_\tau^k) \; .
\end{equation}
Next, we can use $\mu_0^k$ instead of $c_0^k$, and replace the unknown local fractions of the fractions $Y_\gamma$ which are conserved only globally by their local chemical potentials:
\begin{equation}
p^k=p^k(T,\mu_0^k,\mu_\gamma^k,Y_\lambda^k) \; .
\end{equation}
According to eqs.~(\ref{eq_condeq1}) \& (\ref{eq_condeq2}) the remaining chemical potentials have to be equal in the two phases, and the locally fixed fractions, too. Thus:
\begin{equation}
p^I(T,\mu_0,\mu_\gamma,Y_{\lambda})=p^{II}(T,\mu_0,\mu_\gamma,Y_{\lambda}) \; . \label{eq_pexp}
\end{equation}
This formulation, in which the chemical potentials of the globally conserved charges are treated as known state variables, is most convenient to discuss the properties of the phase transition. Of course, the chemical potentials in this equation are actually fixed by the values of the density $c_0$, the fractions $Y_\tau$ and the chosen local constraints.

First we will analyze the case in which no other globally conserved charges besides $C_0$ exist, $G=1$. Eq.~(\ref{eq_pexp}) then leads to a relation
\begin{equation}
\mu_0=\mu^{coex}_0(T, Y_\lambda) \; , \label{eq_mucoex}
\end{equation}
which also fixes the coexistence pressure:
\begin{equation}
p=p^{coex}(T, Y_\lambda) \; .
\end{equation}
This means that there is only one value of the pressure $p^{coex}$ and the chemical potential $\mu^{coex}_0$, where the two phases can coexist. All other local intensive variables are also fixed by the local constraints and the equilibrium conditions and remain constant in the mixed phase, too. A simple way to determine the phase transition pressure is to see where the $p^k(T,\mu_0,Y_\lambda)$-curves of the two phases intersect. The transition from phase I to phase II occurs, when the pressure is equal in the two phases.

The mixed phase will extend over a certain range in the density $c_0$, with the onset given by $c_0^I(T,\mu_0^{coex},Y_\tau)$ and the end by $c_0\2(T,\mu_0^{coex},Y_\lambda)$. Inside the mixed phase, the intensive variables are independent of $c_0$ and $c_0$ is only used to specify the volume fraction $0<X=V\2/V^I<1$ of the two phases:
\begin{equation}
c_0=(1-X)c_0^I(T,\mu_0^{coex},Y_\lambda)+ X c_0\2(T,\mu_0^{coex},Y_\lambda) \; .
\end{equation}
All extensive variables change linearly with the volume fraction in the same way, e.g.:
\begin{equation}
\epsilon_0=(1-X)\epsilon_0^I(T,\mu_0^{coex},Y_\lambda)+ X \epsilon_0\2(T,\mu_0^{coex},Y_\lambda) \; .
\end{equation}
Therefore the calculation of the mixed phase becomes trivial. After the coexistence pressure is found it is given by a linear interpolation in the volume fraction between the onset and endpoint of the mixed phase. This case corresponds to the well known Maxwell construction.

For $G\geq2$ globally conserved charges, eq.~(\ref{eq_pexp}) is not sufficient to determine all chemical potentials $\mu_0, \mu_\gamma$. The equilibrium condition only allows to fix one of the chemical potentials, e.g.~$\mu_0=\mu_0(T, Y_\tau, \mu_{\gamma})$ and $G-1$ chemical potentials remain unknown. But besides the equilibrium conditions, also the total volume and the globally conserved charges have to have the correct value:
\begin{eqnarray}
V&=&V^I+V\2 \nonumber \\
C_\gamma&=&C_\gamma^I(V^I,T,\mu_0,\mu_\gamma,Y_\lambda)+ \nonumber \\
&& C_\gamma\2(V\2,T,\mu_0,\mu_\gamma,Y_\lambda)\; . \label{eq_cgamma}
\end{eqnarray}
These $G+1$ eqs. involve only two further unknowns $V^I$ and $V\2$ so that the whole system of eqs.~(\ref{eq_pexp}) and (\ref{eq_cgamma}) can be solved for given volume $V$, and all thermodynamic variables can be determined. Consequently in this case all quantities (including the pressure) will depend on the values of the densities $c_0$ and $c_\gamma=C_\gamma/V$. A change in the density $c_0$ will also imply a change in the pressure. Thus for $G\geq2$ there will be an extended range in pressure in which the two phases can coexist. The simple Maxwell construction cannot be applied, as the system does not behave linearly any more. Instead it is necessary to calculate the mixed phase at every point $(T,V,C_\sigma)$ explicitly.

In both cases ($G=1$ or $G\geq2$) an extended mixed phase between the two phases forms for the chosen state variables. At the onset of the mixed phase the volume of the newly appearing phase vanishes. Similarly, at the end of the mixed phase only the second phase remains (we assume that the two EoSs of the single phases do not show any discontinuities). Thus the mixed phase becomes identical to the neighboring single phases when approaching the onset or end of the mixed phase. Inside the mixed phase the volume fraction changes continuously from 0 to 1, and consequently all global thermodynamic variables up to first derivatives of the thermodynamic potential will change continuously across the whole phase transition. The second derivatives will in general be discontinuous, as they involve the derivative of the volume fraction.

The thermodynamic potential is the Helmholtz free energy, which also changes continuously and which has the following form inside the mixed phase:
\begin{eqnarray}
F&=&-pV+\sum_{i,k} N_i^k \mu_i^k \nonumber \\
&=&-pV+\mu_{0} C_0 \; ,
\end{eqnarray}
where we used eq.~(\ref{eq_summua}). For $G=1$ in which the pressure and the chemical potentials are constant, the free energy changes linearly with the volume $V$.

Next we want to use the different set of state variables $(T,p, C_\sigma)$, in which the volume is replaced by the pressure and we want to analyze the properties of the phase transition in this new formulation. Now the Gibbs free enthalpy $G=G(T,p,C_\sigma)$, already specified by eq.~(\ref{eq_summua}), is the appropriate thermodynamic potential. These state variables are especially important because they can directly be used for the description of isothermal neutron stars. Under the influence of gravity in a hydrostatic configuration of a compact star, the pressure has to change continuously and has to be strictly monotonic. In the following we only consider a change of the pressure $p$ and keep all the other state variables constant.

If $G=1$ the mixed phase collapses to one single point at the coexistence pressure $p^{coex}$ introduced above. There is only a point of coexistence, but no extended mixed phase. No mixed phase has to be calculated, only the transition point has to be determined. As $\mu_0$ is constant across the mixed phase and continuous at the endpoints, one gets that the potentials of the two phases are equal at the transition point, $G=G^I=G\2$. The equality of $\mu_{0}^I=\mu_0\2$ leads to the equality of the Gibbs free enthalpy.

This equality can also be seen as the reason why the volume fraction of the two phases remains arbitrary at the coexistence point and cannot be determined from the equilibrium conditions. Thus all extensive quantities but those of the externally fixed state variables remain unspecified at the coexistence point.

Because of mechanical, thermal and (at least partial) chemical equilibrium the thermodynamic potential changes continuously across the transition, even though no extended mixed phase exists. For smaller or larger pressures the mixed phase disappears, and only the phase with the lower Gibbs free enthalpy will be present. Despite this, the phase transition is not continuous, as e.g.~the volume behaves discontinuously due to the disappearance of the mixed phase:
\begin{equation}
\lim_{p^< \rightarrow p^{coex}}\left.\frac{\partial G^I}{\partial p}\right|_{T,C_\sigma}=V^I \neq V\2 \lim_{p^> \rightarrow p^{coex}}\left.\frac{\partial G\2}{\partial p}\right|_{T,C_\sigma}  \; .
\end{equation}
Therefore the charge densities, defined by $C_\sigma/V$ will change discontinuously, too. Also the entropy jumps in an analogous way at the phase transition, if $T\neq 0$. The internal energy, given by $E=G-pV+TS$ will also behave discontinuously in general. These discontinuities appear in the first derivatives of the thermodynamic potential. Thus the phase transition is of first order according to the Ehrenfest classification. The discussed scenario is the familiar case of the Maxwell phase transition, e.g.~known from the liquid-gas phase transition of water.

For $G\geq2$ there will be an extended range in pressure in which the two phases can coexist and an extended mixed phase forms. As noted before, the simple Maxwell construction cannot be applied. The mixed phase does not behave linearly any more and thus it has to be calculated explicitly for every single pressure. Now the equilibrium conditions and the know\-ledge of the state variables become sufficient to specify the volume fractions and all other thermodynamic variables of the two phases. This case is usually called the Gibbs construction in the context of cold deleptonized neutron stars with global charge neutrality. The presence of a mixed phases with $X=0$ at the onset and $X=1$ at the endpoint assures that all thermodynamic variables (up to first derivatives) change continuously across the phase transition, as argued above. The phase transition is of second order according to the Ehrenfest classification.

As was shown, locally conserved charge fractions do not influence the qualitative behavior of the phase transition. It is only the number of globally conserved charges which determines the order of the phase transition. Independently of any locally conserved charge fractions, for $G=1$ the system behaves like a simple body. By replacing globally conserved charges by adequate local conservation laws, this allows to reduce the number $G$ of globally conserved charges. In the next section we will use this procedure for isothermal PTs to arrive at a Maxwell construction, even if in principle multiple conserved charges exist.

One can expect that the extension of the mixed phase decreases with the number of local constraints applied. When $C_0$ remains as the only globally conserved charge, the mixed phase will disappear completely inside a compact star. Furthermore the discontinuity of the second derivatives will increase with the number of local constraints. For $G=1$ even the first derivatives become discontinuous then.

The situation becomes different, if we consider a phase transition in an adiabatic instead of an isothermal process. To keep the entropy constant, necessarily the temperature has to change across the transition with equal temperatures in the two phases at each point of the mixed phase. The change in temperature will lead to a change in all other intensive variables, too. Thus even for $G=1$ the Maxwell construction cannot be applied. If mechanical and thermal equilibrium between the two phases is required, the mixed phase does not vanish in a hydrostatic situation under the influence of gravity, even if only one globally conserved charge exists. This is an important result, especially for proto-neutron stars which are very often described by a constant ratio of entropy per baryon as a first approximation. Some other local constraints are needed (e.g.~locally fixed entropy per baryon) which will influence the conditions for phase equilibrium in a non-trivial way. Therefore we will discuss adiabatic phase transition at the end of this article separately.

\section{Application for Matter in Supernovae, Proto-Neutron Stars and Neutron Stars}
\label{sec_app}
As an example, the general relations which were found shall now be applied to the liquid-gas phase transition of nuclear matter and the hadron-quark phase transition under typical astrophysical conditions. The hadronic phases shall consist of $N_\nu$ neutrinos, $N_e$ electrons, $N_p$ protons and $N_n$ neutrons (net numbers, including antiparticles). The two-flavor quark phase shall be composed out of $N_e$ electrons, $N_\nu$ neutrinos, $N_u$ up and $N_d$ down quarks. Furthermore, at the end of the section we will discuss strange quark matter, too, in which $N_s$ strange quarks are also part of the system.

The baryon number $N_B=N_n+N_p$ for hadrons, and $N_B=1/3(N_u+N_d)$ for quarks, and the total electric charge number $N_C=N_p-N_e$, and $N_C=1/3(2N_u-N_d)-N_e$, always have to be conserved in a closed system. Because of charge neutrality $N_C=0$, but the concrete values of the conserved charges are actually irrelevant for the equilibrium conditions of the chemical potentials. Furthermore, there are two additional conserved charges possible, the lepton number $N_L=N_e+N_\nu$ and the proton number $N_p$. The latter in combination with baryon number conservation is equivalent to the conservation of isospin or of baryonic electric charge. Obviously, this is equivalent to flavor conservation for up and down quark matter. Thus, constant $N_p=1/3(2N_u-N_d)$ can be used to express isospin conservation for both kind of phases. To achieve a consistent description of quark and hadronic matter we will use the term proton number or proton fraction conservation as a synonym for the conservation of isospin in the two-flavor quark phase.

Usually instead of fixing $(N_B, N_C, N_L, N_p)$, an intensive formulation in terms of the proton and lepton fractions $Y_p=n_p/n_B$ and $Y_L=(n_e+n_\nu)/n_B$, the baryon number density $n_B$ and charge density $n_C=n_p-n_e=0$ together with a fixed temperature $T$ is used, like e.g.~in Refs.~\cite{lattimer91nourl, shen98,shen98_2}. We will also apply this formulation in the following. Instead of the electric charge per baryon $Y_C=0$ used before we can also take $n_C=0$ to describe the state of the matter. In the thermodynamic limit, the size of the system becomes irrelevant, so that we can assume that the volume $V$ is also known. Obviously then it is completely equivalent to fix $(T, n_B, n_C,Y_p, Y_L, V)$ instead of $(T,N_B,N_C,N_p,N_L,V)$.

Usually for fixed $Y_p$ the neutrinos are not included in the construction of the EoS but are treated separately. The neutrino dynamics play a crucial role in supernovae and proto-neutron stars. To describe the evolution of such systems it is necessary to handle the neutrinos with a detailed dynamical transport scheme in which their emission, scattering and absorption is calculated. We will explain why it is possible to construct the non-neutrino EoS without taking the neutrinos into account, by showing explicitly that the non-neutrino EoS is independent of the neutrino contribution in single phases and in mixed phases, too.

\subsection{Single Homogeneous Phases}
\label{sec_bulk}
\begin{table*}
\begin{center}
\begin{tabular}[b]{c|c|c}
\hline
\hline
 conserved charge  & \multicolumn{2}{c}{chemical potentials}\\
  & hadrons & quarks \\
\hline
\hline
 $N_B^k$ & $\mu_{N_B}^k=(1-Y_p)\mu_n^k+Y_p\mu_p^k$ & $\mu_{N_B}^k=(2-Y_p)\mu_d^k+(1+Y_p)\mu_u^k$\\
 & $+Y_p\mu_e^k+(Y_L-Y_p)\mu_\nu^k$ & $+Y_p\mu_e^k+(Y_L-Y_p)\mu_\nu^k$ \\
\hline
 $Y_C^k$ & $\mu_{N_C}^k=\mu_\nu^k-\mu_e^k$ & $\mu_{N_C}^k=\mu_\nu^k-\mu_e^k$\\
\hline
$Y_p^k$ & $\mu_{N_p}^k=\mu_p^k-\mu_n^k-\mu_\nu^k+\mu_e^k$ & $\mu_{N_p}^k=\mu_u^k-\mu_d^k-\mu_\nu^k+\mu_e^k$ \\
\hline
$Y_L^k$ & $\mu_{N_L}^k=\mu_\nu^k$ & $\mu_{N_L}^k=\mu_\nu^k$ \\
\hline
\hline
\end{tabular}
\end{center}
\caption{The local chemical potentials of the baryon number $N_B$, electric charge $N_C$, proton number $N_p$ and lepton number $N_L$ in terms of the chemical potentials of the particles in one phases if the baryon number and all fractions are fixed locally. The second column is for a hadronic phase composed of neutrons, protons, electrons and neutrinos and the third column for a phase of up and down quarks and electrons and neutrinos. The results also apply for strange quark matter, with $\mu_d=\mu_s$. For global baryon number conservation $\mu_{N_B}^I=\mu_{N_B}\2$ follows. If some of the fractions are conserved only globally and are no longer restricted by local constraints, the corresponding chemical potentials become equal, too: $\mu_\gamma^I=\mu_\gamma\2$.}
\label{table_1}
\end{table*}
In table \ref{table_1} the chemical potentials of the conserved charges $N_B$, $N_C$, $N_p$ and $N_L$ are expressed in terms of the chemical potentials of the particles of one phase if the baryon number and all fractions are kept constant locally, leading to the unusual form of $\mu_{N_B}^k$. As long as only one single homogeneous phase exists, local conservation laws are identical to global ones. The index $k$ can be suppressed if we want to use table \ref{table_1} for one single homogeneous phase.

The special form of the chemical potentials of the conserved charges/fractions can be understood in a simple way. $\mu_{C_\gamma}$ gives the change of the free energy with the change of the charge $C_\gamma$ for constant proton and lepton fraction and electric charge neutrality. The combination of chemical potentials of the particles which is found for $\mu_{C_\gamma}$ corresponds to the change of the particle numbers induced by the change of $C_\gamma$ under the chosen constraints. We note once more, that the form of a chemical potential depends on which other quantities are kept constant. E.g.~the baryon chemical potential $\mu_{N_B}^k$ would be equal to $\mu_n^k$ (for nuclear matter) if instead of the fractions the charges themselves were used as the state variables. However, the final equilibrium conditions are not (and cannot be) affected by the choice of the state variables. Thus we can use the description presented here, which is most convenient for our purpose as it can be applied for single phases as well as for all possible combinations of locally conserved fractions inside mixed phases.

If all the four charges are conserved this corresponds to the situation of matter with completely trapped neutrinos, but too short dynamical timescales to change the proton number by weak reactions. Next we will discuss the different possibilities in which some of the charges are actually not conserved any more. As was shown before, for every charge becoming not conserved an additional equilibrium condition appears. With the new information of eq.~(\ref{eq_mug0g}) then the chemical potentials of the remaining conserved charges can possibly be written in a different simplified form. We note that baryon number and electric charge always have to be conserved.

For non-conserved lepton number from eq.~(\ref{eq_mug0}) and table~\ref{table_1} $\mu_\nu=0$ follows. Neutrinos are completely untrapped/free streaming. If the proton number is still conserved weak reactions involving nucleons are assumed to be completely suppressed. We will discuss the meaning of the result $\mu_\nu=0$ in more detail later.

If neutrinos are completely trapped and there is enough time for the weak reactions to become efficient, the lepton number is conserved but the proton number not. Then the well known weak equilibrium conditions
\begin{equation}
\mu_p-\mu_n-\mu_\nu+\mu_e=0 \label{eq_weakeqh}
\end{equation}
for nuclear matter and
\begin{equation}
\mu_u-\mu_d-\mu_\nu+\mu_e=0 \label{eq_weakeqq}
\end{equation}
for quark matter are found.

If then at a later stage in the evolution the neutrinos become completely untrapped, the lepton number is not conserved any more. Only baryon number and electric charge remain as conserved charges. Then two equilibrium conditions are necessary to derive all particle numbers ${N_i}$. Without lepton number conservation $\mu_\nu=0$ and the neutrinos drop out in the $\beta$-equilibrium conditions
\begin{equation}
\mu_e+\mu_p-\mu_n=0  \;  \label{eq_betaeqh}
\end{equation}
for nucleons and
\begin{equation}
\mu_u-\mu_d-\mu_e=0 \label{eq_betaeqq}
\end{equation}
for quarks. For both sets of particles $\mu_{N_C}=-\mu_e$. The baryon chemical potential can also be expressed in a simpler way: $\mu_{N_B}=\mu_n$ for nucleons and $\mu_{N_B}=2\mu_d+\mu_u$ for quarks.

\subsection{Role of Neutrinos}
\label{ss_nu}
Before we continue, we want to show why neutrinos do not have to be included in the construction of the non-neutrino part of an equation of state in terms of $(T,n_B, Y_p)$, i.e.~if the proton number is conserved.

First we note, that for a single phase $(T,n_B, Y_p)$ are sufficient to fix all particle numbers but neutrinos. The non-neutrino part of the EoS would also not change, if neutrinos were not included as part of the thermodynamic system right from the beginning. At the same time, the neutrino contribution is also independent of the non-neutrino EoS: If the lepton fraction is conserved, i.e.~if they are completely trapped, the neutrino density is directly specified by $n_\nu(T,\mu_\nu)=(Y_L-Y_p)n_B$. Without lepton number conservation, $\mu_\nu=0$ also directly sets the neutrino contribution.

Neutrinos also do not have to be included in the construction of a mixed phase if we only allow global lepton number conservation. Then it always follows that $\mu_\nu^I=\mu_\nu\2$ (the only quantum number of the neutrino is the lepton number so that eq.~(\ref{eq_mumneq2}) applies). In all cases, neutrinos can be treated as ideal Fermi-gases, so that their contribution to the EoS in the two phases is exactly the same. Thus it is also sufficient to study the pressure equilibrium without taking neutrinos into account.

When the lepton fraction is conserved globally, the neutrino chemical potential can also always be taken out of the chemical equilibrium conditions. Because the neutrino chemical potentials and all locally fixed fractions are equal in the two phases, terms proportional to $Y_\lambda \mu_\nu^k$ cancel in the equilibrium conditions of other globally conserved charges. One finds that the same conditions for chemical equilibrium of the non-neutrino part are obtained, as if neutrinos were not present at all. We conclude that neutrinos do not influence the phase equilibrium between the two phases as they are distributed uniformly over the entire system, if the lepton fraction is conserved globally. It is sufficient to calculate an equation of state for protons, neutrons and electrons (or quarks and electrons) in terms of $(T,n_B,Y_p)$ and this equation of state can be used for all possible conditions under which the neutrinos appear.

Instead if $Y_p$ is not conserved, i.e.~one has an EoS in terms of $(T,n_B)$ or $(T,n_B,Y_L)$ the neutrinos influence the rest of the matter via the condition for weak equilibrium. The neutrino contribution needs to be taken into account for the evaluation of the non-neutrino EoS. Thus the non-neutrino EoS will also depend on whether lepton number is conserved or not, because the conditions for weak-equilibrium eqs.~(\ref{eq_weakeqh}) and (\ref{eq_weakeqq}), are different from those for beta-equilibrium, eqs.~(\ref{eq_betaeqh}) and (\ref{eq_betaeqq}). In the former case the lepton number is conserved and the neutrinos are determined by $Y_L$. In the latter case neutrinos are completely untrapped and the neutrino contribution becomes trivial, $\mu_\nu=0$. The same beta-equilibrium conditions eqs.~(\ref{eq_betaeqh}) and (\ref{eq_betaeqq}) are found if neutrinos are not included in the thermodynamic description. In general the EoS depends on the conditions under which the neutrinos appear, as soon as $Y_p$ is not conserved.

Let us now discuss the meaning and some interesting consequences of the result $\mu_\nu=0$ for matter with neutrinos but without lepton number conservation. If a particle $i$ carries no conserved charges ($\alpha_{i\gamma}=0 \; \forall \gamma$) with eq.~(\ref{eq_mui}) one finds immediately that its chemical potential is zero. If neutrinos can be described as an ideal gas in equilibrium it follows that $N_\nu=0$, which means that the number of neutrinos equals the number of antineutrinos. Only if $T=0$ both contributions vanish. Non-conserved $Y_L$ would correspond to the situation when the neutrino mean free path is much larger than the size of the NS so that neutrinos can leave the NS freely. Thus the energy of the system is not conserved, but can be carried away by neutrinos as long as they are abundant. As a logical consequence the NS would cool immediately to $T=0$ if weak reaction rates were fast enough (infinitely large emissivities) to allow to describe the neutrinos as an ideal gas as part of the thermodynamic description. In reality it takes some $10^5$ years until the neutron star has cooled to a core temperature of $\sim 10$ keV and the photon cooling era is reached. The neutrinos are far away from equilibrium, their emissivities have to be calculated, and the description of the cooling process requires detailed numerical simulations \cite{page2006}.

\subsection{Mixed Phases}
If a mixed phase exists, it is crucial whether a charge is conserved globally or locally. In the following we will assume that each of the conserved charges is either conserved globally, or its fraction is conserved locally with equal values in the two phases, as before. Obviously, local constraints in the form $C_\gamma^k=0$ are equivalent to $Y_\gamma^k=0$, i.e.~we can interpret local electric charge neutrality as a locally fixed charge fraction, too. Before we start to discuss all relevant combinations of locally and globally conserved charges, we will analyze the physical meaning of the different local constraints.

\subsubsection*{Local charge neutrality}
Already in \cite{voskresensky03} it was pointed out, that depending on the surface tension and the Debye-screening length, local charge neutrality might be the better approximation for bulk matter. If a large surface tension drives the system to sizes much larger than the Debye screening length, only a negligible small charged surface layer in the order of the Debye screening length remains and the bulk of the matter becomes locally charge neutral. Most calculations for the PT to quark matter indicate that this is indeed the case.

If instead the surface tension is small so that the typical structures are smaller than the Debye screening length, global charge neutrality is the more reasonable assumption. This applies to the liquid-gas PT of nuclear matter, as the Debye screening length is large. Anyhow we include the assumption of local charge neutrality for the nuclear PT in the following discussion, because it is instructive and the corresponding equilibrium conditions can easily be devolved to other kind of phase transitions of nuclear matter.

If one requires that both of the two phases have to be locally charge neutral, two different local electric charge chemical potentials appear in eq.~(\ref{eq_mui}) leading to different chemical potentials of all electrically charged particles (electrons, quarks and protons) in the two phases. If one would do the full calculation including finite-size (surface and Coulomb) effects and without the constraint of local electric charge neutrality, the total chemical potential of charged particles would be shifted by the local electrical potential, e.g. $\tilde \mu_p^I=\mu^I_p+eV^I$ leading to full chemical equilibrium, $\tilde \mu^I_p=\tilde\mu\2_p$, see \cite{voskresensky03}. In the present article we are discussing infinite matter without Coulomb forces, so that the electrical potential cannot be determined and the artificial inequality of the chemical potentials of charged particles cannot be resolved.

\subsubsection*{Locally fixed $Y_p$, $Y_L$ or $n_B$}
There is no physical reason why the proton fraction, the lepton fraction or the baryon number density should be conserved locally or should be equal in the two phases, as there is no long range force between the two phases which is associated with these charges. This would imply that the readjustment of the local proton fraction, lepton fraction and/or baryon density does not take place. Therefore chemical equilibrium with respect to a locally conserved charge is not established between the two phases.

A non-vanishing locally fixed density (e.g.~$n_B^I=n_B\2=n_B$) influences the condition for mechanical equilibrium so that pressure equilibrium is not obtained from the first and second law of thermodynamics any more. For these constraints a change of the subvolumes would imply a change of the local baryon numbers, too. Instead of pressure equilibrium only a combination of the local pressures and the local chemical potentials are equal in the two phases. Consequently, the pressure would change discontinuously at the phase transition. Thus, we will not use constraints of non-vanishing locally fixed densities.

The likewise rather unphysical assumptions of locally fixed $Y_p$ or $Y_L$ might be wanted because they allow to achieve a Maxwell construction of the mixed phase at the cost of only partial chemical equilibrium. Because of the additional conserved charges besides $N_B$ local charge neutrality alone is not sufficient for that (as in the case of cold neutron stars) and at least one other of the conserved charges needs to be fixed locally. This is our motivation to investigate locally fixed $Y_p$ or $Y_L$ and we will focus on this aspect when discussing different scenarios in the following subsections.

\subsubsection*{Discussion of different cases}
In Tables \ref{table_2a} \& \ref{table_2b} all the relevant combinations of local and global conservation laws of the conserved charges for the construction of a mixed phase for isothermal matter in supernovae, proto-NS and cold NS are listed. We assume that the densities and fractions are either conserved globally, or locally in a form $Y_p^I=Y_p\2=Y_p$, $Y_L^I=Y_L\2=Y_L$, $n_C^I=n_C\2=n_C=0$, $n_B^I=n_B\2=n_B$. The final equilibrium conditions are expressed in terms of the chemical potentials of the particles in the two phases, in Table \ref{table_2a} for the liquid-gas phase transition and in Table \ref{table_2b} for the hadron-quark phase transition.

\begin{table*}
\begin{center}
\begin{tabular}[b]{c|c|c|c|c}
\hline
\hline
case & \multicolumn{2}{c|}{conserved densities/fractions} & equilibrium conditions & construction of \\
& globally & locally & & mixed phase\\
\hline
\hline
0 & & $n_B, (Y_p), (Y_L), n_C$ & - & direct\\
\hline
Ia & $n_B$ & $Y_p, Y_L,n_C$ & $(1-Y_p)\mu_n^I+Y_p(\mu_p^I+\mu_e^I)+(Y_L-Y_p)\mu_\nu^I=$ & Maxwell \\
& & & $(1-Y_p)\mu_n\2+Y_p(\mu_p\2+\mu_e\2)+(Y_L-Y_p)\mu_\nu\2$ &\\
\hline
Ib & $n_B$ & $Y_L,n_C$ & $\mu_n^I+Y_L\mu_\nu^I=\mu_n\2+Y_L\mu_\nu\2$ & Maxwell \\
\hline
Ic & $n_B$ & $Y_p,n_C$ & $(1-Y_p)\mu_n^I+Y_p(\mu_p^I+\mu_e^I)=(1-Y_p)\mu_n\2+Y_p(\mu_p\2+\mu_e\2)$  & Maxwell \\
\hline
Id & $n_B$ & $n_C$ & $\mu_n^I=\mu_n\2$ & Maxwell \\
\hline
IIa & $n_B, Y_L$ & $Y_p,n_C$ & $(1-Y_p)\mu_n^I+Y_p(\mu_p^I+\mu_e^I)=(1-Y_p)\mu_n\2+Y_p(\mu_p\2+\mu_e\2)$ & Maxwell/Gibbs \\
&&& $\mu_\nu^I=\mu_\nu\2$ \\
\hline
IIb & $n_B,Y_L$ & $n_C$ & $\mu_n^I=\mu_n\2$ & Gibbs \\
&&& $\mu_\nu^I=\mu_\nu\2$ \\
\hline
IIIa & $n_B, Y_p$ & $Y_L, n_C$ & $\mu_n^I+Y_L\mu_\nu^I=\mu_n\2+Y_L\mu_\nu\2$ & Gibbs \\
&&& $\mu_p^I-\mu_n^I-\mu_\nu^I+\mu_e^I=\mu_p\2-\mu_n\2-\mu_\nu\2+\mu_e\2$ & \\
\hline
IIIb & $n_B,Y_p$ & $n_C$ & $\mu_n^I=\mu_n\2$ & Gibbs \\
&&& $\mu_p^I+\mu_e^I=\mu_p\2+\mu_e\2$ & \\
\hline
IV & $n_B,Y_L, Y_p$ & $n_C$ & $\mu_n^I=\mu_n\2$ & Gibbs \\
&&& $\mu_\nu^I=\mu_\nu\2$ \\
&&& $\mu_p^I+\mu_e^I=\mu_p\2+\mu_e\2$ & \\
\hline
V & $n_B,Y_L, Y_p, n_C$ & & $\mu_n^I=\mu_n\2$ & Gibbs \\
&&& $\mu_\nu^I=\mu_\nu\2$ \\
&&& $\mu_p^I=\mu_p\2$ & \\
&&& $\mu_e^I=\mu_e\2$ & \\
\hline
\hline
\end{tabular}
\end{center}
\caption{Equilibrium conditions for the liquid-gas phase transition of nuclear matter for fixed temperature $T$, baryon number density $n_B$ and charge density $n_C$. The lepton fraction $Y_L$ and proton fraction $Y_p$ are conserved in addition in some cases. These charge densities/fractions are fixed locally (with equal values in the two phases) or globally. If $Y_p$ is not conserved weak equilibrium (eqs.~(\ref{eq_weakeqh}) \& (\ref{eq_weakeqq})) is established in both phases. If $Y_L$ is not conserved $\mu_\nu^I=\mu_\nu\2=0$ is obtained, leading to the same equilibrium conditions as if neutrinos were not included in the thermodynamic system.}
\label{table_2a}
\end{table*}

\begin{table*}
\begin{center}
\begin{tabular}[b]{c|c|c|c|c}
\hline
\hline
case & \multicolumn{2}{c|}{conserved densities/fractions} & equilibrium conditions & construction of \\
& globally & locally & & mixed phase\\
\hline
\hline
0 & & $n_B, (Y_p), (Y_L), n_C$ & - & direct\\
\hline
Ia & $n_B$ & $Y_p, Y_L,n_C$ & $(1-Y_p)\mu_n+Y_p(\mu_p+\mu_e^H)+(Y_L-Y_p)\mu_\nu^H=$ & Maxwell \\
& & & $(2-Y_p)\mu_d+(1+Y_p)\mu_u+Y_p\mu_e^Q+(Y_L-Y_p)\mu_\nu^Q$ &\\
\hline
Ib & $n_B$ & $Y_L,n_C$ & $\mu_n+Y_L\mu_\nu^H=2\mu_d+\mu_u+Y_L\mu_\nu^Q$ & Maxwell \\
\hline
Ic & $n_B$ & $Y_p,n_C$ & $(1-Y_p)\mu_n+Y_p(\mu_p+\mu_e^H)=(2-Y_p)\mu_d+(1+Y_p)\mu_u+Y_p\mu_e^Q$  & Maxwell \\
\hline
Id & $n_B$ & $n_C$ & $\mu_n=2\mu_d+\mu_u$ & Maxwell \\
\hline
IIa & $n_B, Y_L$ & $Y_p,n_C$ & $(1-Y_p)\mu_n+Y_p(\mu_p+\mu_e^H)=(2-Y_p)\mu_d+(1+Y_p)\mu_u+Y_p\mu_e^Q$ & Maxwell/Gibbs \\
&&& $\mu_\nu^H=\mu_\nu^Q$ \\
\hline
IIb & $n_B,Y_L$ & $n_C$ & $\mu_n=2\mu_d+\mu_u$ & Gibbs \\
&&& $\mu_\nu^H=\mu_\nu^Q$ \\
\hline
IIIa & $n_B, Y_p$ & $Y_L, n_C$ & $\mu_n+Y_L\mu_\nu^H=2\mu_d+\mu_u+Y_L\mu_\nu^Q$ & Gibbs \\
&&& $\mu_p-\mu_n-\mu_\nu^H+\mu_e^H=\mu_u-\mu_d-\mu_\nu^Q+\mu_e^Q$ & \\
\hline
IIIb & $n_B,Y_p$ & $n_C$ & $\mu_n=2\mu_d+\mu_u$ & Gibbs \\
&&& $\mu_p+\mu_e^H=2\mu_u+\mu_d+\mu_e^Q$  \\
\hline
IV & $n_B,Y_L, Y_p$ & $n_C$ & $\mu_n=2\mu_d+\mu_u$ & Gibbs \\
&&& $\mu_\nu^H=\mu_\nu^Q$ \\
&&& $\mu_p+\mu_e^H=2\mu_u+\mu_d+\mu_e^Q$\\
\hline
V & $n_B,Y_L, Y_p, n_C$ & & $\mu_n=2\mu_d+\mu_u$ & Gibbs \\
&&& $\mu_\nu^H=\mu_\nu^Q$ \\
&&& $\mu_p=2\mu_u+\mu_d$  \\
&&& $\mu_e^H=\mu_e^Q$ \\
\hline
\hline
\end{tabular}
\end{center}
\caption{As Table \ref{table_2a}, but now for the hadron-quark phase transition. $\mu_d=\mu_s$ is always valid.}
\label{table_2b}
\end{table*}

\subsubsection{Case I}
In case Ia, besides local charge neutrality the proton and lepton fractions are fixed locally and the system has only one globally conserved charge, the baryon number. The only internal degree of freedom of the two phases is the local baryon density. For this case table \ref{table_1} expresses the local chemical potentials belonging to the conserved charges $N_B$, $N_C$, $N_p$ and $N_L$ in terms of the chemical potentials of the particles in the phase.

There is only one global chemical potential with the corresponding equilibrium condition:
\begin{equation}
\mu_{N_B}=\mu_{N_B}^I=\mu_{N_B}\2 \; .
\end{equation}
The condition which is shown in table \ref{table_1} expresses that only combinations of particles can be exchanged which do not change the local proton and lepton fractions and are electrically charge neutral to maintain local charge neutrality. Thus in the liquid gas phase transition only a combination of $1-Y_p$ neutrons, $Y_p$ electrons and protons and $Y_L-Y_p$ neutrinos can be exchanged freely between the two phases. In the hadron-quark phase transition only $(1+Y_p)$ up and $(2-Y_p)$ down quarks, $Y_p$ electrons and $Y_L-Y_p$ neutrinos can be exchanged.

As there is only one globally conserved charge $N_B$, the pressure is constant across the phase transition. The Maxwell construction can be used and all the results for $G=1$ of section \ref{sec_proppt} can be applied. The chemical potentials of all particles are different in the two phases, as all particles contribute to the locally conserved fractions.

We note that this case was already discussed in Ref.~\cite{lugones98}. Equivalent equilibrium conditions were found and the same conclusions about the disappearance of the mixed phase in a compact star were drawn.

In the following we will use the results of case Ia in table \ref{table_1} to derive the equilibrium conditions for all other cases. When a fraction is not conserved locally but only globally, the two local chemical potentials specify the new equilibrium conditions. In eq.~(\ref{eq_condeq2}) it was deduced that by conserving $Y_\gamma$ instead of $Y_\gamma^k$, the two local chemical potentials $\mu_\gamma^k$ become equal. If one of the charges is actually not conserved any more, this has to be seen as a global criterion. To minimize the free energy with respect to this charge, in eq.~(\ref{eq_mug0}) it was derived that a non-conserved fraction leads to two new local constraints for the chemical potentials, $\mu_{\delta}^I=\mu_{\delta}\2=0$, which replace the two locally fixed fractions used before.

In case Ib $Y_p$ is no longer conserved. By setting $\mu_{N_p}^k=0$ the weak equilibrium conditions (\ref{eq_weakeqh}) respectively (\ref{eq_weakeqq}) of section \ref{sec_bulk} are obtained which now have to be fulfilled in both phases. In Ic the conservation of $Y_L$ is lifted, leading to $\mu_\nu^I=\mu_\nu\2=0$. If both fractions are not conserved any more as in case Id, the beta-equilibrium conditions (\ref{eq_betaeqh}) respectively (\ref{eq_betaeqq}) are obtained. These results are independent of the other local or global conservation laws. Thus we do not need to discuss the non-conservation of lepton and/or proton fraction in the following cases again.

In cases Ia to Id the different conserved charges allow to rewrite the equilibrium condition $\mu_{N_B}^I=\mu_{N_B}\2$ in the simplified forms presented in tables \ref{table_2a} \& \ref{table_2b}. Case Id describes a cold, deleptonized neutron star. Only global baryon number conservation and local charge neutrality are considered. The well-known result of the equality of the neutron chemical potentials is found for the Maxwell construction of the liquid-gas phase transition. In all other Maxwell constructions the equality of the baryon chemical potential $\mu_{N_B}^k$ takes a different form and involves additional particles besides the neutrons. Because of the inequality of $\mu_{Y_C}^k$ in the two phases the chemical potentials of the electrically charged particles always remain different in the two phases in all cases Ia to Id.

\subsubsection{Case II}
In case II lepton number and baryon number are conserved globally. The second equilibrium condition $\mu_{N_L}^I=\mu_{N_L}\2$ from the global conservation of lepton number leads to the equality of the neutrino chemical potentials. Neutrinos are the only particles which can be exchanged between the two phases, if the baryon number, the electric charge and the proton fraction are kept constant in both phases.

Case IIa assumes locally fixed $Y_p$ and local charge neutrality in addition. The same equilibrium condition as in case Ic in which $Y_L$ is not conserved is obtained for the non-neutrino part of the EoS. Case Ic gives the same description of the non-neutrino EoS as case IIa. If one does not include the neutrinos in the thermodynamic description at all, the same condition as in Ic are found. Once more this shows explicitly that the non-neutrino EoS is independent of the neutrino contribution. As discussed before, the neutrinos can be calculated separately as long as $Y_p$ is conserved and $Y_L$ is not fixed locally. From this point of view, the non-neutrino EoSs of cases with fixed $Y_p$ and globally conserved $Y_L$ are equivalent to fixed $Y_p$ and non-conserved $Y_L$.

For the non-neutrino EoS $G=1$, and the mixed phase without neutrinos can be calculated with the Maxwell construction. The mixed phase would disappear under the influence of gravity in a hydrostatic configuration. But the inclusion of neutrinos leads to an interesting effect on the mixed phase: The neutrino contribution is simply given by $n_\nu(T,\mu_\nu)=(Y_L-Y_p)n_B$. Thus for increasing baryon density also the neutrino density has to increase. Therefore the neutrino pressure is not constant across the phase transition, which is in agreement with our general result for $G=2$.  If the pressure is used as the continuously varying variable and is changed strictly monotonic (e.g.~in a compact star), a mixed phase appears only because of the presence of neutrinos.

It is very interesting to see, that all cases with local charge neutrality in which $Y_p$ or $Y_L$ are conserved globally will lead to an extended mixed phase in a compact star. After the star has cooled to $T=0$ and became completely deleptonized, $Y_p$ and $Y_L$ are no longer conserved, and case Id will be reached. Consequently the mixed phase will disappear during the evolution of the star. We note that such a scenario has not been considered in the literature so far.

If we compare case IIa to case Ia, we see that the same fractions and charges are conserved. In both cases only the Maxwell construction is needed, but in case IIa the unphysical assumption of a locally fixed lepton fraction is not used. Thus case IIa should be preferred over case Ia.

We conclude that case IIa (or equivalently Ic for the non-neutrino EoS) is the most convenient scenario which leads for fixed $Y_p$ to the desired Maxwell construction of the system without neutrinos. All other cases with conserved proton fraction $Y_p$ involve more than one globally conserved charge for the non-neutrino EoS and the explicit evaluation of phase equilibrium is necessary. Because of the additional global conservation of the proton fraction, even with local charge neutrality a simple Maxwell construction is not possible for matter in supernovae or proto-neutron stars.

In case IIb $Y_p$ is no longer conserved, so that the separation of the neutrino EoS is not possible. The Gibbs construction has to be done with the inclusion of neutrinos. Case IIb is physically meaningful, as local charge neutrality is the only local conservation law.

\subsubsection{Case III}
The proton fraction is conserved globally in case III. The general equilibrium condition $\mu_{N_p}^I=\mu_{N_p}\2$ shown by case IIIa in tables \ref{table_2a} \& \ref{table_2b} expresses that only a proton and an electron can be moved from one phase into the other, if at the same time a neutron and a neutrino are converted backwards. All other combinations of particles would change the local baryon number, the electric charge or the lepton fraction.

In case IIIa the neutrino EoS cannot be separated from the rest of the EoS, as the lepton fraction is conserved locally so that the Gibbs construction has to be performed. If one is only interested to achieve the Maxwellian case (without further reasoning why locally fixed $Y_L$ instead of locally fixed $Y_p$ is assumed), the easier case IIa can be applied instead.

Local charge neutrality is the only local conservation law in case IIIb, which is the proper description of a phase transition with sufficiently large surface tension between the two phases. The equilibrium conditions for the local baryon chemical potentials simplify compared to Ic, in which the proton fraction was fixed locally.

\subsubsection{Case IV}
The non-neutrino constraints of case IV are equivalent to those of case IIIb. Case IV gives the correct description of locally charge neutral matter with neutrinos as part of the thermodynamic description.

\subsubsection{Case V}
In case V local electric charge neutrality is not required any more, so that all charges are conserved globally. Eq.~(\ref{eq_mumneq2}) applies now for all particles in the liquid-gas phase transition, and the chemical potentials of all particles become equal. For the hadron-quark phase transition the hadronic chemical potentials directly fix the quark chemical potentials and vice versa.

$Y_L$ and $Y_p$ are conserved globally in cases IV and V. Thus the equilibrium conditions remain the same if one or both of the fractions are no longer conserved. But every non-conserved fraction gives rise to two new stronger local constraints. They contain the information about the chemical equilibrium between the two phases with respect to this fraction used before, so that one of the equilibrium conditions in tables \ref{table_2a} \& \ref{table_2b} becomes meaningless.

\subsubsection{Case 0}
Finally, we want to discuss case 0 in which the phase transition is somewhat constructed by hand. In case 0, all state variables are fixed locally. No mixed phase has to be calculated, as also the baryon density is fixed locally: $n_B^I=n_B\2=n_B$. If conserved at all, the three conserved fractions are fixed locally, too. In case 0 there are no globally conserved charges so that no chemical equilibrium condition between the two phases is obtained. Thus, the two equations of state of the two phases can be calculated completely separately and the phase transition point is then set by one freely selectable condition.

Similarly to the Maxwell construction for the state variable $p$, the subvolumes of the two phases remain arbitrary at the transition point and are not constrained by the equilibrium conditions. Accordingly the extensive variables cannot be determined, too. On the other hand, the local intensive variables remain independent of the volumes of the two phases. Thus the chemical potentials of the locally conserved charges, the local pressure and the local temperature remain well-defined even without knowing the two subvolumes. The two phases can be treated as independent single homogeneous phases with unknown volume.

If pressure equilibrium is taken as the criterion for the determination of the phase transition point (which in this case is not a consequence of the first and second law of thermodynamics any more) the two phases can only coexist at one special density $n_B^{coex}$ if the other state variables are kept constant. No extended mixed phase appears, and all thermodynamic quantities but the pressure and the chemical potentials of electrons for conserved $Y_p$ and neutrinos for conserved $Y_p$ and $Y_L$ change discontinuously across the phase transition. If electrons and neutrinos can be treated as ideal gases $\mu_e^I=\mu_e\2$ follows from the conservation of $Y_p$ and $\mu_\nu^I=\mu_\nu\2$ from the conservation of $Y_p$ and $Y_L$. If instead of the baryon number density the pressure is used as the continuous variable no mixed phase forms, either. At the transition point pressure and thermal equilibrium are established, but at least chemical equilibrium of the baryons is not. Most importantly, therefore the thermodynamic potential, the free energy $F=-pV+\sum_i N_i \mu_i$, behaves discontinuously. Because of the local constraints, the sum $\sum_i \mu_i N_i$ will not be equal in the two phases. If the pressure is used instead of the volume as one of the state variables, the free enthalpy will also behave discontinuously when the transition point is crossed. Thus, the thermodynamic potential can also not be used to determine which of the two phases exists before and which one after the phase transition. The second law of thermodynamics is violated and thus some other additional criterion has to be considered for that.

Besides pressure equilibrium every other possible coexistence condition can be applied. The conclusions remain the same, and in general all the conditions will lead to a discontinuous thermodynamic potential.

Instead one can use the continuity of the corresponding thermodynamic potential (the free energy $F$ for $(T,V,C_\gamma)$, the free enthalpy $G$ for $(T,p,C_\gamma)$) as the phase coexistence criterion. At the point where the two potentials are equal the phase transition occurs. Before and after the phase transition only the phase with the lower potential is present. However, with this choice the pressure (and all other thermodynamic quantities but the state variables) will behave discontinuously.

We note that case 0 with fixed $Y_p$ would imply for the liquid-gas phase transition, that actually no phase transition occurs, because the same particles appear with equal densities in the two phases. Case 0 is only relevant if some internal degrees of freedom remain which can be different in the two phases.

\subsection{EoS in $(S/N_B,p,Y_L,n_C)$}
Apart from numerical simulations, the EoS is usually calculated for the state variables $(S/N_B,p,Y_L,n_C)$. Such an EoS can be used directly for the description of typical representative configurations of proto-neutron stars which are characterized by a constant lepton to baryon ratio in a first approximation.

Because the pressure is chosen to be the independent continuously changing variable, there exist only two possibilities: Either the phase transition occurs only at one single pressure $p^{coex}$, or a mixed phase of the two phases forms over an extended range in pressure. In the first case no mixed phase needs to be calculated. If thermal and mechanical equilibrium are required, the direct phase transition point can be found where the temperatures of the two phases become equal $T^I=T\2$, with locally fixed $S/N_B^I=S/N_B\2=S/N_B$, $p^I=p\2=p$, $Y_L^I=Y_L\2=Y_L$, $n_C^I=n_C\2=n_C$ and (arbitrary) globally conserved baryon number $N_B$ which can be shared by the two phases. Pressure equilibrium is automatically given, as the pressure is one of the state variables which is set to equal values in the two phases.

This case is the adiabatic equivalence to case 0 of the isothermal phase transitions, in which all state variables but the volume were fixed locally. Here temperature equilibrium is chosen as the constraint which determines the coexistence point. Even though thermal equilibrium is enforced, the thermodynamic potential, which is the enthalpy $H=TS+\sum_i N_i \mu_i$, will change discontinuously at the transition point. In contrast in the Maxwell construction of an isothermal phase transition the coexistence pressure is determined by the proper equilibrium conditions of the chemical potentials. We showed that this is equivalent to a continuous thermodynamic potential.

There exists nothing similar for an adiabatic process to the isothermal case IIa, which allows an easy construction of the non-neutrino mixed phase but leads to a second order phase transition with the inclusion of neutrinos. As explained before, the non-neutrino part of the EoS depends on the neutrino fraction as soon as $Y_p$ is not conserved any more. Thus it is not possible to construct the non-neutrino mixed phase by means of an equal pressure Maxwell construction and treat the neutrinos independently. In the adiabatic case, in all scenarios (except for the direct phase transitions) the Gibbs construction has to be used and the mixed phase has to be calculated explicitly with the contribution of the neutrinos.

The only physical meaningful local constraint is local charge neutrality, corresponding to the case of a large surface energy between the two phases. Globally conserved $(S/N_B,Y_L)$ is equivalent to globally conserved $(S,N_B,N_L)$ for constant (arbitrary) baryon number $N_B$. Thus the conditions for chemical equilibrium are the same as in case IIb. As $Y_p$ is not conserved, weak equilibrium is established in the two phases. Only neutrinos and neutrons can be exchanged independently between the two phases in the liquid-gas phase transition. For the hadron-quark phase transition two down
 and one up quark can be converted into a neutron.

The possibility of a locally charge neutral extended mixed phase does not exist for cold deleptonized neutron stars. So far such a kind of a mixed phase was not studied in the literature. We will analyze the properties of this new mixed phase and examine its disappearance during the cooling of the neutron star in a forthcoming separate study \cite{pagliara09}.

If local charge neutrality is lifted, no local charges exist any more, and chemical equilibrium is expressed by case V, with weak equilibrium in the two phases in addition. For the liquid-gas phase transition the chemical potentials of all particles are equal in the two phases.

Because of the stronger constraint of local charge neutrality, one can expect that the locally charge neutral mixed phase will extend over a smaller range in pressure. In both cases the appearance of the mixed phase assures that all thermodynamic variables are well defined and the first derivatives of the thermodynamic potential change continuously with the pressure.

\subsection{Strange Matter}
In strange quark matter, in addition to the up and down quarks, $N_s$ strange quarks are part of the thermodynamic system. In principle, strange quarks carry the additional quantum number of strangeness. There exist two possibilities to handle this additional quantum number: First, one can use the total strangeness of the system indeed as an additional conserved charge. If strangeness is not taken to be identical to zero, it is necessary to calculate the EoS for all possible strangeness fractions $Y_S=N_S/N_B$. The strangeness chemical potential $\mu_{N_S}$ would appear in addition to the chemical potentials of the other conserved charges. This approach was e.g.~used in Ref.~\cite{greiner87} to describe strangeness separation in heavy ion collisions. In this early work for the first time the phase transition to strange quark matter was described by two separate conserved charges, baryon number and strangeness, leading to the phase transition of a multicomponent body. Here we will not discuss the scenario of conserved strangeness any further but will leave it for future discussion.

Second, there exists a simpler and more commonly used description of strange matter, by assuming equilibrium with respect to strangeness changing reactions, i.e.~strangeness is not conserved, so that $\mu_{N_S}=0$. In this case the quantum numbers of the strange quark are identical to the ones of the down quark, so that $\mu_d=\mu_s$ because of eq.~(\ref{eq_mumneq1}). Then all results presented for two-flavor quark matter in this article can also be applied to strange quark matter, with the baryon number given by $N_B=1/3(N_u+N_d+N_s)$ and the electric charge number by $N_C=1/3(2N_u-N_d-N_s)-N_e$.

The only subtlety arises when $Y_p$ is conserved. First of all it is necessary to reconsider the meaning of $Y_p$ for strange matter. One possibility would be to interpret $Y_p$ as the net electric charge carried by baryons, $N_p=1/3(2N_u-N_d-N_s)$, so that $N_p=N_e$ still gives charge neutrality. In combination with baryon number conservation, the conservation of $Y_p$ leads then to a fixed number of up quarks, but only the sum of down and strange quarks is fixed, i.e.~reactions which change down into strange quarks are still in equilibrium. Thus it would be necessary to assume that these reactions happen on a much shorter timescale than reactions which change the number of up quarks (semi-leptonic reactions).

The argumentation followed in Ref.~\cite{sagert09} is a different one. The EoS of state is calculated for fixed $Y_p$. But within the application in a core-collapse supernova, quark matter appears only at such large densities and temperatures, that neutrinos are completely trapped and weak equilibrium is established. Hence $Y_p$ is actually not conserved but only $Y_L$ remains approximately constant. Within the numerical simulation for a given $Y_L$ the proper $Y_p$ is determined. Under these conditions, weak equilibrium is a reasonable assumption and it is not necessary to argument with the different timescales of the different reactions.

The whole discussion of this subsection applies also for hyperonic matter, i.e.~hadronic matter with strangeness, and the conclusions are analog.

\section{Summary \& Conclusions}
\label{sec_sum}
In section \ref{sec_eq} we derived the chemical equilibrium conditions for the chemical potentials of the particles. We used a formulation, in which the two local parts $C_0^k$ of the two phases $k=I, II$ of the conserved charge $C_0=C_0^I+C_0\2$ and the local charge fractions $Y_\tau^k=C_\tau^k/C_0^k$ of the conserved charges $C_\tau=C_\tau^I+C_\tau\2$ were chosen to be the independent degrees of freedom. Finally in eqs.~(\ref{eq_mui}), the chemical potentials of the particles were expressed by the chemical potentials of these degrees of freedom. We selected this special set of state variables because it can be used for all additional local constraints which are considered in this article: locally fixed fractions with equal values in the two phases, or local charge densities which are zero (e.g.~local charge neutrality). From this general formulation we continued with a particular set of local and global conservation laws and showed that the local chemical potentials of globally conserved charges without local constraints become equal, see eqs.~(\ref{eq_condeq1}) \& (\ref{eq_condeq2}). I.e.~if no local constraint exists for a charge, chemical equilibrium with respect to this charge is established. In eq.~(\ref{eq_summua}) it became apparent, that $\mu_0 C_0$ (where $\mu_0$ is the chemical potential of the charge $C_0$) is equivalent to the Gibbs potential.

With these results we showed in section \ref{sec_proppt} that the qualitative properties of the phase transition only depend on the number of globally conserved charges $G$ and that they are independent of locally fixed fractions. In case there is only one globally conserved charge, $G=1$, the simple Maxwell construction can be used for the calculation of the mixed phase. The mixed phase will vanish in a static configuration under the influence of gravity, leading to a discontinuous phase transition with a continuous thermodynamic potential. If there is more than one globally conserved charge, $G \geq 2$, the Gibbs construction applies and every point inside the mixed phase has to be calculated explicitly. An extended mixed phase always forms, leading to a second order phase transition. The assumption of additional local constraints allows to achieve the Maxwell construction, if all the charge fractions are fixed locally. However, this does not work for an adiabatic phase transition, as the temperature has to vary with the density to keep the entropy constant.

We applied these general findings to phase transitions in typical astrophysical systems in section \ref{sec_app}. We considered the liquid-gas phase transition and the transition from hadronic to quark matter at high densities. It was shown that the results for up and down quark matter are the same as for strange quark matter with $\mu_d=\mu_s$ if equilibrium with respect to strangeness changing reactions is assumed. First we focused on single phases, and derived that the equilibrium condition with a non-conserved fraction is given by setting the corresponding chemical potential to zero. The well known weak equilibrium conditions (eqs.~(\ref{eq_weakeqh}) \& (\ref{eq_weakeqq})) are found if the proton fraction is not conserved, and the conditions for beta equilibrium (eqs.~(\ref{eq_betaeqh}) \& (\ref{eq_betaeqq})) if both the lepton and the proton fraction are not conserved. If the lepton fraction is not conserved $\mu_\nu^I=\mu_\nu\2=0$, if neutrinos are included in the thermodynamic system. Furthermore it was shown that the neutrino EoS is independent of the rest of the matter as long as $Y_p$ is conserved and $Y_L$ is not fixed locally.

For mixed phases it is crucial whether a local or a global conservation law is applied. We argued that local charge neutrality might give a better description of mixed phases, if the typical size of the structures is larger than the Debye screening length, due to a large surface tension between the two phases. Instead there is no physical motivation for a locally fixed proton or lepton fraction, as no long range force is associated with these charges. However, we used these assumptions to reduce the number of globally conserved charges.

We classified the equilibrium conditions between the two phases for all relevant cases (see tables \ref{table_2a} \& \ref{table_2b}). All of them were derived by applying the procedure described in section \ref{sec_eq}: If a charge is conserved only globally its chemical potentials become equal in the two phases. For this we used the results from table \ref{table_1} which expresses the chemical potentials of the conserved charges in terms of the chemical potentials of the particles, if all of the fractions are fixed locally. If $Y_p$ or $Y_p$ and $Y_L$ are no longer conserved, this is a global constraint leading to weak- or beta-equilibrium in both of the two phases. The case Id, in which $Y_p$ and $Y_L$ are not conserved, gives the well known Maxwell construction for cold-deleptonized neutron stars.

Next, we discussed all the different cases. For fixed $Y_p$ case IIa was most interesting, as it allowed to use the Maxwell construction for the non-neutrino EoS with locally fixed $Y_p$ as the only additional local constraint in addition to local charge neutrality. With the inclusion of neutrinos and globally fixed $Y_L$, an extended mixed phase will form, causing a second order phase transition. The non-neutrino part of case IIa was equivalent to case Ic. Compared to case Ia, case IIa should be used preferably, as the Maxwell construction is achieved without the assumption of a locally fixed lepton fraction. Case IIIa is also based on a locally fixed lepton fraction, but still requires the Gibbs construction. If the Maxwell construction is not necessary, case IIIb and IV show the equilibrium conditions for local charge neutrality. In case V no local constraints are applied, resulting in the equality of the chemical potentials of identical particles in the two phases.

For non-conserved $Y_p$ nothing analogous to case IIa does exist, as the non-neutrino part becomes dependent on the neutrino contribution. Either one assumes a locally fixed lepton fraction (case Ib) or the Gibbs construction has to be used. The only physical meaningful local constraint in the latter case is local charge neutrality (case IIb). The same is true for an adiabatic phase transition in which no Maxwell construction can be achieved at all.

The possibility of a direct phase transition always exists, by fixing all the state variables locally so that no mixed phase has to be calculated at all. However this is contradictory to the second law of thermodynamics as the thermodynamic potential changes discontinuously.

It would be interesting to study the quantitative properties of the mixed phase and the phase transition by applying the general conditions found in the present article to specific equations of state for hadronic and quark matter and to analyze the consequences for the proto-neutron stars' evolution during the cooling process and its stability. Another promising investigation would be the implications of the different assumptions for the mixed phase in the dynamical environment of a supernova, as done e.g.~in \cite{sagert09}.

The unphysical assumptions of locally fixed fractions have to be seen as a tool to construct a mixed phase in a thermodynamic consistent way, without the need of complicated calculations of the mixed phase. So far, the necessary conditions for the Maxwell construction were not derived explicitly in the literature.

The assumption of local charge neutrality is instead physical meaningful and represents a simple way to simulate strong Coulomb forces, without the need to include finite-size and Coulomb effects. Yet, local charge neutral mixed phases have not been analyzed in a lepton-rich, hot environment for the quark-hadron phase transition. Such locally charge neutral structures can grow almost arbitrary in size. One can expect significant changes of dynamical properties like the neutrino emissivities and opacities or the thermal conductivity. Furthermore local charge neutrality has an interesting effect on the evolution of the mixed phase itself. As long as $Y_p$ or $Y_L$ are conserved at least two globally conserved charges exist and a mixed phase is present. After the deleptonization, finally only $N_B$ will remain as a globally conserved charge, and when the star becomes isothermal the mixed phase will disappear. The authors are currently preparing a separate study \cite{pagliara09} in which these effects will be analyzed for some specific EoSs.

There is another application of the results in this article, namely the description of the nucleation of the quark phase. As the initially formed bubble will have a small size, surface effects become crucial. The deconfinement process is mediated only by strong interactions. Since the timescales for any weak reactions are much longer, one can assume that this leads to flavor conservation during the deconfinement transition. Depending on the size of the relevant fluctuations in the hadronic phase one might choose one of the different cases presented in table \ref{table_2b} to calculate the threshold density of the nucleation. E.g.~in \cite{lugones98} the equilibrium conditions denoted by case Ia in the present article were derived with the motivation to determine the phase transition point under the assumption of flavor conservation. The authors assume locally fixed $Y_p$ and $Y_L$ and local charge neutrality. However, in this calculation chemical equilibrium with respect to baryon number is still established, leading to different baryon baryon densities in the two phases, although the proton fraction remains equal in the two phases. One might consider the even stronger constraint, that initially also the baryon density has to remain constant inside the bubble which is nucleated. After the quarks have been deconfined, weak reactions will take place and the system will develop to a state described by case IV for local charge neutrality or by case V for global charge neutrality. Latent heat will be released during this non-equilibrium process.

The present work may also be relevant for the description of the QCD phase transition in relativistic heavy ion collisions, explored in experiments at e.g.~CERN, RHIC or FAIR. If the relativistic expansion of the fireball in the central collision zone is characterized by a constant entropy and baryon number, from our considerations we can conclude that a mixed phase exists over an extended range in pressure and thus over a finite period in time, too. The phase transition will be continuous, if thermodynamic equilibrium is reached at all stages of the expansion. The description of matter in the QCD phase transition as a multi-component system allows different choices of the equilibrium conditions, as presented here in the case of compact stars. E.g.~in Ref.~\cite{lonanno07} in addition to global conservation of baryon number the global conservation of isospin was considered. The recent investigation \cite{steinheimer09} indicates, that also strangeness fluctuations and separation could lead to interesting effects. However, in contrast to the case of compact stars, the typical hydrocells in a heavy ion collision are usually very small of order $0.1$ fm. Thus it might be questionable to use an EoS which includes a mixed phase for the hydrodynamic description. We note that only such globally conserved charges are relevant degrees of freedom, which are actually accessible by the system.

\subsection*{Acknowledgments}

M.H.~acknowledges support from the Graduate Program for Hadron and Ion Research (GP-HIR). The work of G.P.~is supported by the Alliance Program of the Helmholtz Association (HA216/EMMI). J.S.B.~is supported by the German Research Foundation (DFG) within the framework of the excellence initiative through the Heidelberg Graduate School of Fundamental Physics. This work has been supported by CompStar a research networking programme of the European Science Foundation. We would like to thank Irina Sagert for her important contribution to this project. 

\bibliographystyle{apsrev}
\bibliography{literat}

\begin{thebibliography}{54}
\expandafter\ifx\csname natexlab\endcsname\relax\def\natexlab#1{#1}\fi
\expandafter\ifx\csname bibnamefont\endcsname\relax
  \def\bibnamefont#1{#1}\fi
\expandafter\ifx\csname bibfnamefont\endcsname\relax
  \def\bibfnamefont#1{#1}\fi
\expandafter\ifx\csname citenamefont\endcsname\relax
  \def\citenamefont#1{#1}\fi
\expandafter\ifx\csname url\endcsname\relax
  \def\url#1{\texttt{#1}}\fi
\expandafter\ifx\csname urlprefix\endcsname\relax\def\urlprefix{URL }\fi
\providecommand{\bibinfo}[2]{#2}
\providecommand{\eprint}[2][]{\url{#2}}

\bibitem[{\citenamefont{{Ravenhall} et~al.}(1983)\citenamefont{{Ravenhall},
  {Pethick}, and {Wilson}}}]{ravenhall83}
\bibinfo{author}{\bibfnamefont{D.~G.} \bibnamefont{{Ravenhall}}},
  \bibinfo{author}{\bibfnamefont{C.~J.} \bibnamefont{{Pethick}}},
  \bibnamefont{and} \bibinfo{author}{\bibfnamefont{J.~R.}
  \bibnamefont{{Wilson}}}, \bibinfo{journal}{Phys. Rev. Lett.}
  \textbf{\bibinfo{volume}{50}}, \bibinfo{pages}{2066} (\bibinfo{year}{1983}).

\bibitem[{\citenamefont{{Lamb} et~al.}(1983)\citenamefont{{Lamb}, {Lattimer},
  {Pethick}, and {Ravenhall}}}]{lamb83}
\bibinfo{author}{\bibfnamefont{D.~Q.} \bibnamefont{{Lamb}}},
  \bibinfo{author}{\bibfnamefont{J.~M.} \bibnamefont{{Lattimer}}},
  \bibinfo{author}{\bibfnamefont{C.~J.} \bibnamefont{{Pethick}}},
  \bibnamefont{and} \bibinfo{author}{\bibfnamefont{D.~G.}
  \bibnamefont{{Ravenhall}}}, \bibinfo{journal}{Nuclear Physics A}
  \textbf{\bibinfo{volume}{411}}, \bibinfo{pages}{449} (\bibinfo{year}{1983}).

\bibitem[{\citenamefont{{M{\"u}ller} and {Serot}}(1995)}]{muller95}
\bibinfo{author}{\bibfnamefont{H.}~\bibnamefont{{M{\"u}ller}}}
  \bibnamefont{and} \bibinfo{author}{\bibfnamefont{B.~D.}
  \bibnamefont{{Serot}}}, \bibinfo{journal}{\prc}
  \textbf{\bibinfo{volume}{52}}, \bibinfo{pages}{2072} (\bibinfo{year}{1995}),
  \eprint{arXiv:nucl-th/9505013}.

\bibitem[{\citenamefont{{Ishizuka} et~al.}(2003)\citenamefont{{Ishizuka},
  {Ohnishi}, and {Sumiyoshi}}}]{ishizuka03}
\bibinfo{author}{\bibfnamefont{C.}~\bibnamefont{{Ishizuka}}},
  \bibinfo{author}{\bibfnamefont{A.}~\bibnamefont{{Ohnishi}}},
  \bibnamefont{and}
  \bibinfo{author}{\bibfnamefont{K.}~\bibnamefont{{Sumiyoshi}}},
  \bibinfo{journal}{Nucl. Phys. A} \textbf{\bibinfo{volume}{723}},
  \bibinfo{pages}{517} (\bibinfo{year}{2003}), \eprint{arXiv:nucl-th/0208020}.

\bibitem[{\citenamefont{{Bugaev} et~al.}(2001)\citenamefont{{Bugaev},
  {Gorenstein}, {Mishustin}, and {Greiner}}}]{bugaev01}
\bibinfo{author}{\bibfnamefont{K.~A.} \bibnamefont{{Bugaev}}},
  \bibinfo{author}{\bibfnamefont{M.~I.} \bibnamefont{{Gorenstein}}},
  \bibinfo{author}{\bibfnamefont{I.~N.} \bibnamefont{{Mishustin}}},
  \bibnamefont{and}
  \bibinfo{author}{\bibfnamefont{W.}~\bibnamefont{{Greiner}}},
  \bibinfo{journal}{Physics Letters B} \textbf{\bibinfo{volume}{498}},
  \bibinfo{pages}{144} (\bibinfo{year}{2001}), \eprint{arXiv:nucl-th/0011055}.

\bibitem[{\citenamefont{{Itoh}}(1970)}]{itoh70}
\bibinfo{author}{\bibfnamefont{N.}~\bibnamefont{{Itoh}}},
  \bibinfo{journal}{Progress of Theoretical Physics}
  \textbf{\bibinfo{volume}{44}}, \bibinfo{pages}{291} (\bibinfo{year}{1970}).

\bibitem[{\citenamefont{{Heiselberg} et~al.}(1993)\citenamefont{{Heiselberg},
  {Pethick}, and {Staubo}}}]{heiselberg93}
\bibinfo{author}{\bibfnamefont{H.}~\bibnamefont{{Heiselberg}}},
  \bibinfo{author}{\bibfnamefont{C.~J.} \bibnamefont{{Pethick}}},
  \bibnamefont{and} \bibinfo{author}{\bibfnamefont{E.~F.}
  \bibnamefont{{Staubo}}}, \bibinfo{journal}{Physical Review Letters}
  \textbf{\bibinfo{volume}{70}}, \bibinfo{pages}{1355} (\bibinfo{year}{1993}).

\bibitem[{\citenamefont{{Glendenning} and
  {Schaffner-Bielich}}(1998)}]{glendenning98}
\bibinfo{author}{\bibfnamefont{N.~K.} \bibnamefont{{Glendenning}}}
  \bibnamefont{and}
  \bibinfo{author}{\bibfnamefont{J.}~\bibnamefont{{Schaffner-Bielich}}},
  \bibinfo{journal}{Physical Review Letters} \textbf{\bibinfo{volume}{81}},
  \bibinfo{pages}{4564} (\bibinfo{year}{1998}),
  \eprint{arXiv:astro-ph/9810284}.

\bibitem[{\citenamefont{{Glendenning} and
  {Schaffner-Bielich}}(1999)}]{glendenning99}
\bibinfo{author}{\bibfnamefont{N.~K.} \bibnamefont{{Glendenning}}}
  \bibnamefont{and}
  \bibinfo{author}{\bibfnamefont{J.}~\bibnamefont{{Schaffner-Bielich}}},
  \bibinfo{journal}{\prc} \textbf{\bibinfo{volume}{60}},
  \bibinfo{pages}{025803} (\bibinfo{year}{1999}),
  \eprint{arXiv:astro-ph/9810290}.

\bibitem[{\citenamefont{{Fujii} et~al.}(1996)\citenamefont{{Fujii}, {Maruyama},
  {Muto}, and {Tatsumi}}}]{fujii96}
\bibinfo{author}{\bibfnamefont{H.}~\bibnamefont{{Fujii}}},
  \bibinfo{author}{\bibfnamefont{T.}~\bibnamefont{{Maruyama}}},
  \bibinfo{author}{\bibfnamefont{T.}~\bibnamefont{{Muto}}}, \bibnamefont{and}
  \bibinfo{author}{\bibfnamefont{T.}~\bibnamefont{{Tatsumi}}},
  \bibinfo{journal}{Nuclear Physics A} \textbf{\bibinfo{volume}{597}},
  \bibinfo{pages}{645} (\bibinfo{year}{1996}).

\bibitem[{\citenamefont{{Pons} et~al.}(2000)\citenamefont{{Pons}, {Reddy},
  {Ellis}, {Prakash}, and {Lattimer}}}]{pons00}
\bibinfo{author}{\bibfnamefont{J.~A.} \bibnamefont{{Pons}}},
  \bibinfo{author}{\bibfnamefont{S.}~\bibnamefont{{Reddy}}},
  \bibinfo{author}{\bibfnamefont{P.~J.} \bibnamefont{{Ellis}}},
  \bibinfo{author}{\bibfnamefont{M.}~\bibnamefont{{Prakash}}},
  \bibnamefont{and} \bibinfo{author}{\bibfnamefont{J.~M.}
  \bibnamefont{{Lattimer}}}, \bibinfo{journal}{\prc}
  \textbf{\bibinfo{volume}{62}}, \bibinfo{pages}{035803}
  (\bibinfo{year}{2000}), \eprint{arXiv:nucl-th/0003008}.

\bibitem[{\citenamefont{{Haensel} and {Proszynski}}(1982)}]{haensel82}
\bibinfo{author}{\bibfnamefont{P.}~\bibnamefont{{Haensel}}} \bibnamefont{and}
  \bibinfo{author}{\bibfnamefont{M.}~\bibnamefont{{Proszynski}}},
  \bibinfo{journal}{\apj} \textbf{\bibinfo{volume}{258}}, \bibinfo{pages}{306}
  (\bibinfo{year}{1982}).

\bibitem[{\citenamefont{{Migdal} et~al.}(1990)\citenamefont{{Migdal},
  {Saperstein}, {Troitsky}, and {Voskresensky}}}]{migdal90}
\bibinfo{author}{\bibfnamefont{A.~B.} \bibnamefont{{Migdal}}},
  \bibinfo{author}{\bibfnamefont{E.~E.} \bibnamefont{{Saperstein}}},
  \bibinfo{author}{\bibfnamefont{M.~A.} \bibnamefont{{Troitsky}}},
  \bibnamefont{and} \bibinfo{author}{\bibfnamefont{D.~N.}
  \bibnamefont{{Voskresensky}}}, \bibinfo{journal}{Phys. Rep.}
  \textbf{\bibinfo{volume}{192}}, \bibinfo{pages}{179} (\bibinfo{year}{1990}).

\bibitem[{\citenamefont{{Migdal} et~al.}(1979)\citenamefont{{Migdal},
  {Chernoutsan}, and {Mishustin}}}]{migdal79}
\bibinfo{author}{\bibfnamefont{A.~B.} \bibnamefont{{Migdal}}},
  \bibinfo{author}{\bibfnamefont{A.~I.} \bibnamefont{{Chernoutsan}}},
  \bibnamefont{and} \bibinfo{author}{\bibfnamefont{I.~N.}
  \bibnamefont{{Mishustin}}}, \bibinfo{journal}{Physics Letters B}
  \textbf{\bibinfo{volume}{83}}, \bibinfo{pages}{158} (\bibinfo{year}{1979}).

\bibitem[{\citenamefont{{Schaffner-Bielich}
  et~al.}(2002)\citenamefont{{Schaffner-Bielich}, {Hanauske}, {St{\"o}cker},
  and {Greiner}}}]{schaffner02}
\bibinfo{author}{\bibfnamefont{J.}~\bibnamefont{{Schaffner-Bielich}}},
  \bibinfo{author}{\bibfnamefont{M.}~\bibnamefont{{Hanauske}}},
  \bibinfo{author}{\bibfnamefont{H.}~\bibnamefont{{St{\"o}cker}}},
  \bibnamefont{and}
  \bibinfo{author}{\bibfnamefont{W.}~\bibnamefont{{Greiner}}},
  \bibinfo{journal}{Physical Review Letters} \textbf{\bibinfo{volume}{89}},
  \bibinfo{pages}{171101} (\bibinfo{year}{2002}),
  \eprint{arXiv:astro-ph/0005490}.

\bibitem[{\citenamefont{{R{\"u}ster} et~al.}(2006)\citenamefont{{R{\"u}ster},
  {Werth}, {Buballa}, {Shovkovy}, and {Rischke}}}]{Ruster06nourl}
\bibinfo{author}{\bibfnamefont{S.~B.} \bibnamefont{{R{\"u}ster}}},
  \bibinfo{author}{\bibfnamefont{V.}~\bibnamefont{{Werth}}},
  \bibinfo{author}{\bibfnamefont{M.}~\bibnamefont{{Buballa}}},
  \bibinfo{author}{\bibfnamefont{I.~A.} \bibnamefont{{Shovkovy}}},
  \bibnamefont{and} \bibinfo{author}{\bibfnamefont{D.~H.}
  \bibnamefont{{Rischke}}}, \bibinfo{journal}{Phys. Rev. D}
  \textbf{\bibinfo{volume}{73}}, \bibinfo{pages}{034025}
  (\bibinfo{year}{2006}).

\bibitem[{\citenamefont{{Blaschke} et~al.}(2005)\citenamefont{{Blaschke},
  {Fredriksson}, {Grigorian}, {{\"O}zta{\c s}}, and {Sandin}}}]{blaschke06}
\bibinfo{author}{\bibfnamefont{D.}~\bibnamefont{{Blaschke}}},
  \bibinfo{author}{\bibfnamefont{S.}~\bibnamefont{{Fredriksson}}},
  \bibinfo{author}{\bibfnamefont{H.}~\bibnamefont{{Grigorian}}},
  \bibinfo{author}{\bibfnamefont{A.~M.} \bibnamefont{{{\"O}zta{\c s}}}},
  \bibnamefont{and} \bibinfo{author}{\bibfnamefont{F.}~\bibnamefont{{Sandin}}},
  \bibinfo{journal}{\prd} \textbf{\bibinfo{volume}{72}},
  \bibinfo{pages}{065020} (\bibinfo{year}{2005}),
  \eprint{arXiv:hep-ph/0503194}.

\bibitem[{\citenamefont{{Pagliara} and {Schaffner-Bielich}}(2008)}]{pagliara08}
\bibinfo{author}{\bibfnamefont{G.}~\bibnamefont{{Pagliara}}} \bibnamefont{and}
  \bibinfo{author}{\bibfnamefont{J.}~\bibnamefont{{Schaffner-Bielich}}},
  \bibinfo{journal}{\prd} \textbf{\bibinfo{volume}{77}},
  \bibinfo{pages}{063004} (\bibinfo{year}{2008}), \eprint{arXiv:0711.1119}.

\bibitem[{\citenamefont{{Fryer} and {Woosley}}(1998)}]{fryer98}
\bibinfo{author}{\bibfnamefont{C.~L.} \bibnamefont{{Fryer}}} \bibnamefont{and}
  \bibinfo{author}{\bibfnamefont{S.~E.} \bibnamefont{{Woosley}}},
  \bibinfo{journal}{Astrophys. J.} \textbf{\bibinfo{volume}{501}},
  \bibinfo{pages}{780} (\bibinfo{year}{1998}).

\bibitem[{\citenamefont{{Drago} et~al.}(2008)\citenamefont{{Drago}, {Pagliara},
  and {Schaffner-Bielich}}}]{drago08}
\bibinfo{author}{\bibfnamefont{A.}~\bibnamefont{{Drago}}},
  \bibinfo{author}{\bibfnamefont{G.}~\bibnamefont{{Pagliara}}},
  \bibnamefont{and}
  \bibinfo{author}{\bibfnamefont{J.}~\bibnamefont{{Schaffner-Bielich}}},
  \bibinfo{journal}{Journal of Physics G Nuclear Physics}
  \textbf{\bibinfo{volume}{35}}, \bibinfo{pages}{014052}
  (\bibinfo{year}{2008}), \eprint{arXiv:0705.4418}.

\bibitem[{\citenamefont{{Berezhiani} et~al.}(2003)\citenamefont{{Berezhiani},
  {Bombaci}, {Drago}, {Frontera}, and {Lavagno}}}]{berezhiani03}
\bibinfo{author}{\bibfnamefont{Z.}~\bibnamefont{{Berezhiani}}},
  \bibinfo{author}{\bibfnamefont{I.}~\bibnamefont{{Bombaci}}},
  \bibinfo{author}{\bibfnamefont{A.}~\bibnamefont{{Drago}}},
  \bibinfo{author}{\bibfnamefont{F.}~\bibnamefont{{Frontera}}},
  \bibnamefont{and}
  \bibinfo{author}{\bibfnamefont{A.}~\bibnamefont{{Lavagno}}},
  \bibinfo{journal}{\apj} \textbf{\bibinfo{volume}{586}}, \bibinfo{pages}{1250}
  (\bibinfo{year}{2003}), \eprint{arXiv:astro-ph/0209257}.

\bibitem[{\citenamefont{{Mishustin} et~al.}(2003)\citenamefont{{Mishustin},
  {Hanauske}, {Bhattacharyya}, {Satarov}, {St{\"o}cker}, and
  {Greiner}}}]{mishustin03}
\bibinfo{author}{\bibfnamefont{I.~N.} \bibnamefont{{Mishustin}}},
  \bibinfo{author}{\bibfnamefont{M.}~\bibnamefont{{Hanauske}}},
  \bibinfo{author}{\bibfnamefont{A.}~\bibnamefont{{Bhattacharyya}}},
  \bibinfo{author}{\bibfnamefont{L.~M.} \bibnamefont{{Satarov}}},
  \bibinfo{author}{\bibfnamefont{H.}~\bibnamefont{{St{\"o}cker}}},
  \bibnamefont{and}
  \bibinfo{author}{\bibfnamefont{W.}~\bibnamefont{{Greiner}}},
  \bibinfo{journal}{Physics Letters B} \textbf{\bibinfo{volume}{552}},
  \bibinfo{pages}{1} (\bibinfo{year}{2003}), \eprint{arXiv:hep-ph/0210422}.

\bibitem[{\citenamefont{{Glendenning} et~al.}(1997)\citenamefont{{Glendenning},
  {Pei}, and {Weber}}}]{glendenning97}
\bibinfo{author}{\bibfnamefont{N.~K.} \bibnamefont{{Glendenning}}},
  \bibinfo{author}{\bibfnamefont{S.}~\bibnamefont{{Pei}}}, \bibnamefont{and}
  \bibinfo{author}{\bibfnamefont{F.}~\bibnamefont{{Weber}}},
  \bibinfo{journal}{Physical Review Letters} \textbf{\bibinfo{volume}{79}},
  \bibinfo{pages}{1603} (\bibinfo{year}{1997}),
  \eprint{arXiv:astro-ph/9705235}.

\bibitem[{\citenamefont{{Zdunik} et~al.}(2006)\citenamefont{{Zdunik}, {Bejger},
  {Haensel}, and {Gourgoulhon}}}]{zdunik06}
\bibinfo{author}{\bibfnamefont{J.~L.} \bibnamefont{{Zdunik}}},
  \bibinfo{author}{\bibfnamefont{M.}~\bibnamefont{{Bejger}}},
  \bibinfo{author}{\bibfnamefont{P.}~\bibnamefont{{Haensel}}},
  \bibnamefont{and}
  \bibinfo{author}{\bibfnamefont{E.}~\bibnamefont{{Gourgoulhon}}},
  \bibinfo{journal}{Astron. Astrophys.} \textbf{\bibinfo{volume}{450}},
  \bibinfo{pages}{747} (\bibinfo{year}{2006}), \eprint{arXiv:astro-ph/0509806}.

\bibitem[{\citenamefont{{Reddy} et~al.}(2000)\citenamefont{{Reddy}, {Bertsch},
  and {Prakash}}}]{reddy00}
\bibinfo{author}{\bibfnamefont{S.}~\bibnamefont{{Reddy}}},
  \bibinfo{author}{\bibfnamefont{G.}~\bibnamefont{{Bertsch}}},
  \bibnamefont{and}
  \bibinfo{author}{\bibfnamefont{M.}~\bibnamefont{{Prakash}}},
  \bibinfo{journal}{Physics Letters B} \textbf{\bibinfo{volume}{475}},
  \bibinfo{pages}{1} (\bibinfo{year}{2000}), \eprint{arXiv:nucl-th/9909040}.

\bibitem[{\citenamefont{{Glendenning}}(2001)}]{glendenning01}
\bibinfo{author}{\bibfnamefont{N.~K.} \bibnamefont{{Glendenning}}},
  \bibinfo{journal}{Phys. Rep.} \textbf{\bibinfo{volume}{342}},
  \bibinfo{pages}{393} (\bibinfo{year}{2001}).

\bibitem[{\citenamefont{{Blaschke} et~al.}(2001)\citenamefont{{Blaschke},
  {Grigorian}, and {Poghosyan}}}]{blaschke01}
\bibinfo{author}{\bibfnamefont{D.}~\bibnamefont{{Blaschke}}},
  \bibinfo{author}{\bibfnamefont{H.}~\bibnamefont{{Grigorian}}},
  \bibnamefont{and}
  \bibinfo{author}{\bibfnamefont{G.}~\bibnamefont{{Poghosyan}}}, in
  \emph{\bibinfo{booktitle}{Physics of Neutron Star Interiors}}, edited by
  \bibinfo{editor}{\bibfnamefont{D.}~\bibnamefont{{Blaschke}}},
  \bibinfo{editor}{\bibfnamefont{N.~K.} \bibnamefont{{Glendenning}}},
  \bibnamefont{and}
  \bibinfo{editor}{\bibfnamefont{A.}~\bibnamefont{{Sedrakian}}}
  (\bibinfo{year}{2001}), vol. \bibinfo{volume}{578} of
  \emph{\bibinfo{series}{Lecture Notes in Physics, Berlin Springer Verlag}}, p.
  \bibinfo{pages}{285}.

\bibitem[{\citenamefont{{Page} et~al.}(2006)\citenamefont{{Page}, {Geppert},
  and {Weber}}}]{page2006}
\bibinfo{author}{\bibfnamefont{D.}~\bibnamefont{{Page}}},
  \bibinfo{author}{\bibfnamefont{U.}~\bibnamefont{{Geppert}}},
  \bibnamefont{and} \bibinfo{author}{\bibfnamefont{F.}~\bibnamefont{{Weber}}},
  \bibinfo{journal}{Nucl. Phys. A} \textbf{\bibinfo{volume}{777}},
  \bibinfo{pages}{497} (\bibinfo{year}{2006}), \eprint{arXiv:astro-ph/0508056}.

\bibitem[{\citenamefont{{Glendenning}}(1992)}]{glendenning92}
\bibinfo{author}{\bibfnamefont{N.~K.} \bibnamefont{{Glendenning}}},
  \bibinfo{journal}{\prd} \textbf{\bibinfo{volume}{46}}, \bibinfo{pages}{1274}
  (\bibinfo{year}{1992}).

\bibitem[{\citenamefont{{M{\"u}ller}}(1997)}]{muller97}
\bibinfo{author}{\bibfnamefont{H.}~\bibnamefont{{M{\"u}ller}}},
  \bibinfo{journal}{Nuclear Physics A} \textbf{\bibinfo{volume}{618}},
  \bibinfo{pages}{349} (\bibinfo{year}{1997}), \eprint{arXiv:nucl-th/9701035}.

\bibitem[{\citenamefont{{Voskresensky}
  et~al.}(2003)\citenamefont{{Voskresensky}, {Yasuhira}, and
  {Tatsumi}}}]{voskresensky03}
\bibinfo{author}{\bibfnamefont{D.~N.} \bibnamefont{{Voskresensky}}},
  \bibinfo{author}{\bibfnamefont{M.}~\bibnamefont{{Yasuhira}}},
  \bibnamefont{and}
  \bibinfo{author}{\bibfnamefont{T.}~\bibnamefont{{Tatsumi}}},
  \bibinfo{journal}{Nuclear Physics A} \textbf{\bibinfo{volume}{723}},
  \bibinfo{pages}{291} (\bibinfo{year}{2003}), \eprint{arXiv:nucl-th/0208067}.

\bibitem[{\citenamefont{{Maruyama} et~al.}(2007)\citenamefont{{Maruyama},
  {Chiba}, {Schulze}, and {Tatsumi}}}]{maruyama07}
\bibinfo{author}{\bibfnamefont{T.}~\bibnamefont{{Maruyama}}},
  \bibinfo{author}{\bibfnamefont{S.}~\bibnamefont{{Chiba}}},
  \bibinfo{author}{\bibfnamefont{H.-J.} \bibnamefont{{Schulze}}},
  \bibnamefont{and}
  \bibinfo{author}{\bibfnamefont{T.}~\bibnamefont{{Tatsumi}}},
  \bibinfo{journal}{\prd} \textbf{\bibinfo{volume}{76}},
  \bibinfo{pages}{123015} (\bibinfo{year}{2007}), \eprint{arXiv:0708.3277}.

\bibitem[{\citenamefont{{Maruyama}
  et~al.}(2008{\natexlab{a}})\citenamefont{{Maruyama}, {Chiba}, {Schulze}, and
  {Tatsumi}}}]{maruyama08a}
\bibinfo{author}{\bibfnamefont{T.}~\bibnamefont{{Maruyama}}},
  \bibinfo{author}{\bibfnamefont{S.}~\bibnamefont{{Chiba}}},
  \bibinfo{author}{\bibfnamefont{H.-J.} \bibnamefont{{Schulze}}},
  \bibnamefont{and}
  \bibinfo{author}{\bibfnamefont{T.}~\bibnamefont{{Tatsumi}}},
  \bibinfo{journal}{Physics Letters B} \textbf{\bibinfo{volume}{659}},
  \bibinfo{pages}{192} (\bibinfo{year}{2008}{\natexlab{a}}),
  \eprint{arXiv:nucl-th/0702088}.

\bibitem[{\citenamefont{{Maruyama}
  et~al.}(2008{\natexlab{b}})\citenamefont{{Maruyama}, {Chiba}, {Schulze}, and
  {Tatsumi}}}]{maruyama08b}
\bibinfo{author}{\bibfnamefont{T.}~\bibnamefont{{Maruyama}}},
  \bibinfo{author}{\bibfnamefont{S.}~\bibnamefont{{Chiba}}},
  \bibinfo{author}{\bibfnamefont{H.-J.} \bibnamefont{{Schulze}}},
  \bibnamefont{and}
  \bibinfo{author}{\bibfnamefont{T.}~\bibnamefont{{Tatsumi}}},
  \bibinfo{journal}{Journal of Physics G Nuclear Physics}
  \textbf{\bibinfo{volume}{35}}, \bibinfo{pages}{104076}
  (\bibinfo{year}{2008}{\natexlab{b}}), \eprint{arXiv:0711.3501}.

\bibitem[{\citenamefont{{Pons} et~al.}(2001)\citenamefont{{Pons}, {Steiner},
  {Prakash}, and {Lattimer}}}]{pons01}
\bibinfo{author}{\bibfnamefont{J.~A.} \bibnamefont{{Pons}}},
  \bibinfo{author}{\bibfnamefont{A.~W.} \bibnamefont{{Steiner}}},
  \bibinfo{author}{\bibfnamefont{M.}~\bibnamefont{{Prakash}}},
  \bibnamefont{and} \bibinfo{author}{\bibfnamefont{J.~M.}
  \bibnamefont{{Lattimer}}}, \bibinfo{journal}{Physical Review Letters}
  \textbf{\bibinfo{volume}{86}}, \bibinfo{pages}{5223} (\bibinfo{year}{2001}),
  \eprint{arXiv:astro-ph/0102015}.

\bibitem[{\citenamefont{{Maruyama} et~al.}(2006)\citenamefont{{Maruyama},
  {Tatsumi}, {Voskresensky}, {Tanigawa}, {Endo}, and {Chiba}}}]{maruyama06}
\bibinfo{author}{\bibfnamefont{T.}~\bibnamefont{{Maruyama}}},
  \bibinfo{author}{\bibfnamefont{T.}~\bibnamefont{{Tatsumi}}},
  \bibinfo{author}{\bibfnamefont{D.~N.} \bibnamefont{{Voskresensky}}},
  \bibinfo{author}{\bibfnamefont{T.}~\bibnamefont{{Tanigawa}}},
  \bibinfo{author}{\bibfnamefont{T.}~\bibnamefont{{Endo}}}, \bibnamefont{and}
  \bibinfo{author}{\bibfnamefont{S.}~\bibnamefont{{Chiba}}},
  \bibinfo{journal}{\prc} \textbf{\bibinfo{volume}{73}},
  \bibinfo{pages}{035802} (\bibinfo{year}{2006}),
  \eprint{arXiv:nucl-th/0505063}.

\bibitem[{\citenamefont{{Maruyama} et~al.}(2005)\citenamefont{{Maruyama},
  {Tatsumi}, {Voskresensky}, {Tanigawa}, and {Chiba}}}]{maruyama05}
\bibinfo{author}{\bibfnamefont{T.}~\bibnamefont{{Maruyama}}},
  \bibinfo{author}{\bibfnamefont{T.}~\bibnamefont{{Tatsumi}}},
  \bibinfo{author}{\bibfnamefont{D.~N.} \bibnamefont{{Voskresensky}}},
  \bibinfo{author}{\bibfnamefont{T.}~\bibnamefont{{Tanigawa}}},
  \bibnamefont{and} \bibinfo{author}{\bibfnamefont{S.}~\bibnamefont{{Chiba}}},
  \bibinfo{journal}{\prc} \textbf{\bibinfo{volume}{72}},
  \bibinfo{pages}{015802} (\bibinfo{year}{2005}),
  \eprint{arXiv:nucl-th/0503027}.

\bibitem[{\citenamefont{{Sagert} et~al.}(2009)\citenamefont{{Sagert},
  {Fischer}, {Hempel}, {Pagliara}, {Schaffner-Bielich}, {Mezzacappa},
  {Thielemann}, and {Liebend{\"o}rfer}}}]{sagert09}
\bibinfo{author}{\bibfnamefont{I.}~\bibnamefont{{Sagert}}},
  \bibinfo{author}{\bibfnamefont{T.}~\bibnamefont{{Fischer}}},
  \bibinfo{author}{\bibfnamefont{M.}~\bibnamefont{{Hempel}}},
  \bibinfo{author}{\bibfnamefont{G.}~\bibnamefont{{Pagliara}}},
  \bibinfo{author}{\bibfnamefont{J.}~\bibnamefont{{Schaffner-Bielich}}},
  \bibinfo{author}{\bibfnamefont{A.}~\bibnamefont{{Mezzacappa}}},
  \bibinfo{author}{\bibfnamefont{F.-K.} \bibnamefont{{Thielemann}}},
  \bibnamefont{and}
  \bibinfo{author}{\bibfnamefont{M.}~\bibnamefont{{Liebend{\"o}rfer}}},
  \bibinfo{journal}{Physical Review Letters} \textbf{\bibinfo{volume}{102}},
  \bibinfo{pages}{081101} (\bibinfo{year}{2009}), \eprint{arXiv:0809.4225}.

\bibitem[{\citenamefont{{Nakazato} et~al.}(2008)\citenamefont{{Nakazato},
  {Sumiyoshi}, and {Yamada}}}]{nakazato08}
\bibinfo{author}{\bibfnamefont{K.}~\bibnamefont{{Nakazato}}},
  \bibinfo{author}{\bibfnamefont{K.}~\bibnamefont{{Sumiyoshi}}},
  \bibnamefont{and} \bibinfo{author}{\bibfnamefont{S.}~\bibnamefont{{Yamada}}},
  \bibinfo{journal}{Phys. Rev. D} \textbf{\bibinfo{volume}{77}},
  \bibinfo{pages}{103006} (\bibinfo{year}{2008}), \eprint{arXiv:0804.0661}.

\bibitem[{\citenamefont{Bauswein et~al.}(2009)\citenamefont{Bauswein, Janka,
  Oechslin, Pagliara, Sagert, Schaffner-Bielich, Hohle, and
  Neuh\"{a}user}}]{bauswein09}
\bibinfo{author}{\bibfnamefont{A.}~\bibnamefont{Bauswein}},
  \bibinfo{author}{\bibfnamefont{H.-T.} \bibnamefont{Janka}},
  \bibinfo{author}{\bibfnamefont{R.}~\bibnamefont{Oechslin}},
  \bibinfo{author}{\bibfnamefont{G.}~\bibnamefont{Pagliara}},
  \bibinfo{author}{\bibfnamefont{I.}~\bibnamefont{Sagert}},
  \bibinfo{author}{\bibfnamefont{J.}~\bibnamefont{Schaffner-Bielich}},
  \bibinfo{author}{\bibfnamefont{M.~M.} \bibnamefont{Hohle}}, \bibnamefont{and}
  \bibinfo{author}{\bibfnamefont{R.}~\bibnamefont{Neuh\"{a}user}},
  \bibinfo{journal}{Physical Review Letters} \textbf{\bibinfo{volume}{103}},
  \bibinfo{eid}{011101} (\bibinfo{year}{2009}).

\bibitem[{\citenamefont{{Prakash} et~al.}(1997)\citenamefont{{Prakash},
  {Bombaci}, {Prakash}, {Ellis}, {Lattimer}, and {Knorren}}}]{prakash97}
\bibinfo{author}{\bibfnamefont{M.}~\bibnamefont{{Prakash}}},
  \bibinfo{author}{\bibfnamefont{I.}~\bibnamefont{{Bombaci}}},
  \bibinfo{author}{\bibfnamefont{M.}~\bibnamefont{{Prakash}}},
  \bibinfo{author}{\bibfnamefont{P.~J.} \bibnamefont{{Ellis}}},
  \bibinfo{author}{\bibfnamefont{J.~M.} \bibnamefont{{Lattimer}}},
  \bibnamefont{and}
  \bibinfo{author}{\bibfnamefont{R.}~\bibnamefont{{Knorren}}},
  \bibinfo{journal}{Phys. Rep.} \textbf{\bibinfo{volume}{280}},
  \bibinfo{pages}{1} (\bibinfo{year}{1997}), \eprint{arXiv:nucl-th/9603042}.

\bibitem[{\citenamefont{{Prakash} et~al.}(1995)\citenamefont{{Prakash},
  {Cooke}, and {Lattimer}}}]{prakash95}
\bibinfo{author}{\bibfnamefont{M.}~\bibnamefont{{Prakash}}},
  \bibinfo{author}{\bibfnamefont{J.~R.} \bibnamefont{{Cooke}}},
  \bibnamefont{and} \bibinfo{author}{\bibfnamefont{J.~M.}
  \bibnamefont{{Lattimer}}}, \bibinfo{journal}{\prd}
  \textbf{\bibinfo{volume}{52}}, \bibinfo{pages}{661} (\bibinfo{year}{1995}).

\bibitem[{\citenamefont{{Burgio} and {Plumari}}(2008)}]{burgio08}
\bibinfo{author}{\bibfnamefont{G.~F.} \bibnamefont{{Burgio}}} \bibnamefont{and}
  \bibinfo{author}{\bibfnamefont{S.}~\bibnamefont{{Plumari}}},
  \bibinfo{journal}{\prd} \textbf{\bibinfo{volume}{77}},
  \bibinfo{pages}{085022} (\bibinfo{year}{2008}), \eprint{arXiv:0710.5384}.

\bibitem[{\citenamefont{{Yasuhira} and {Tatsumi}}(2001)}]{yasuhira01}
\bibinfo{author}{\bibfnamefont{M.}~\bibnamefont{{Yasuhira}}} \bibnamefont{and}
  \bibinfo{author}{\bibfnamefont{T.}~\bibnamefont{{Tatsumi}}},
  \bibinfo{journal}{Nuclear Physics A} \textbf{\bibinfo{volume}{690}},
  \bibinfo{pages}{769} (\bibinfo{year}{2001}), \eprint{arXiv:nucl-th/0009090}.

\bibitem[{\citenamefont{{Nicotra} et~al.}(2006)\citenamefont{{Nicotra},
  {Baldo}, {Burgio}, and {Schulze}}}]{nicotra06}
\bibinfo{author}{\bibfnamefont{O.~E.} \bibnamefont{{Nicotra}}},
  \bibinfo{author}{\bibfnamefont{M.}~\bibnamefont{{Baldo}}},
  \bibinfo{author}{\bibfnamefont{G.~F.} \bibnamefont{{Burgio}}},
  \bibnamefont{and} \bibinfo{author}{\bibfnamefont{H.-J.}
  \bibnamefont{{Schulze}}}, \bibinfo{journal}{\prd}
  \textbf{\bibinfo{volume}{74}}, \bibinfo{pages}{123001}
  (\bibinfo{year}{2006}), \eprint{arXiv:astro-ph/0608021}.

\bibitem[{\citenamefont{{Yasutake} and {Kashiwa}}(2009)}]{yasutake09}
\bibinfo{author}{\bibfnamefont{N.}~\bibnamefont{{Yasutake}}} \bibnamefont{and}
  \bibinfo{author}{\bibfnamefont{K.}~\bibnamefont{{Kashiwa}}},
  \bibinfo{journal}{\prd} \textbf{\bibinfo{volume}{79}},
  \bibinfo{pages}{043012} (\bibinfo{year}{2009}), \eprint{arXiv:0902.0111}.

\bibitem[{\citenamefont{{Lugones} and {Benvenuto}}(1998)}]{lugones98}
\bibinfo{author}{\bibfnamefont{G.}~\bibnamefont{{Lugones}}} \bibnamefont{and}
  \bibinfo{author}{\bibfnamefont{O.~G.} \bibnamefont{{Benvenuto}}},
  \bibinfo{journal}{\prd} \textbf{\bibinfo{volume}{58}},
  \bibinfo{pages}{083001} (\bibinfo{year}{1998}).

\bibitem[{\citenamefont{{Shen} et~al.}(1998{\natexlab{a}})\citenamefont{{Shen},
  {Toki}, {Oyamatsu}, and {Sumiyoshi}}}]{shen98}
\bibinfo{author}{\bibfnamefont{H.}~\bibnamefont{{Shen}}},
  \bibinfo{author}{\bibfnamefont{H.}~\bibnamefont{{Toki}}},
  \bibinfo{author}{\bibfnamefont{K.}~\bibnamefont{{Oyamatsu}}},
  \bibnamefont{and}
  \bibinfo{author}{\bibfnamefont{K.}~\bibnamefont{{Sumiyoshi}}},
  \bibinfo{journal}{Nucl. Phys. A} \textbf{\bibinfo{volume}{637}},
  \bibinfo{pages}{435} (\bibinfo{year}{1998}{\natexlab{a}}),
  \eprint{arXiv:nucl-th/9805035}.

\bibitem[{\citenamefont{{Shen} et~al.}(1998{\natexlab{b}})\citenamefont{{Shen},
  {Toki}, {Oyamatsu}, and {Sumiyoshi}}}]{shen98_2}
\bibinfo{author}{\bibfnamefont{H.}~\bibnamefont{{Shen}}},
  \bibinfo{author}{\bibfnamefont{H.}~\bibnamefont{{Toki}}},
  \bibinfo{author}{\bibfnamefont{K.}~\bibnamefont{{Oyamatsu}}},
  \bibnamefont{and}
  \bibinfo{author}{\bibfnamefont{K.}~\bibnamefont{{Sumiyoshi}}},
  \bibinfo{journal}{Progress of Theoretical Physics}
  \textbf{\bibinfo{volume}{100}}, \bibinfo{pages}{1013}
  (\bibinfo{year}{1998}{\natexlab{b}}), \eprint{arXiv:nucl-th/9806095}.

\bibitem[{\citenamefont{{Lattimer} and {Douglas
  Swesty}}(1991)}]{lattimer91nourl}
\bibinfo{author}{\bibfnamefont{J.~M.} \bibnamefont{{Lattimer}}}
  \bibnamefont{and} \bibinfo{author}{\bibfnamefont{F.}~\bibnamefont{{Douglas
  Swesty}}}, \bibinfo{journal}{Nucl. Phys. A} \textbf{\bibinfo{volume}{535}},
  \bibinfo{pages}{331} (\bibinfo{year}{1991}).

\bibitem[{\citenamefont{{Pagliara} et~al.}(2009)\citenamefont{{Pagliara},
  {Hempel}, and {Schaffner-Bielich}}}]{pagliara09}
\bibinfo{author}{\bibfnamefont{G.}~\bibnamefont{{Pagliara}}},
  \bibinfo{author}{\bibfnamefont{M.}~\bibnamefont{{Hempel}}}, \bibnamefont{and}
  \bibinfo{author}{\bibfnamefont{J.}~\bibnamefont{{Schaffner-Bielich}}},
  \bibinfo{journal}{{in preparation}}  (\bibinfo{year}{2009}).

\bibitem[{\citenamefont{{Greiner} et~al.}(1987)\citenamefont{{Greiner}, {Koch},
  and {Stocker}}}]{greiner87}
\bibinfo{author}{\bibfnamefont{C.}~\bibnamefont{{Greiner}}},
  \bibinfo{author}{\bibfnamefont{P.}~\bibnamefont{{Koch}}}, \bibnamefont{and}
  \bibinfo{author}{\bibfnamefont{H.}~\bibnamefont{{Stocker}}},
  \bibinfo{journal}{Physical Review Letters} \textbf{\bibinfo{volume}{58}},
  \bibinfo{pages}{1825} (\bibinfo{year}{1987}).

\bibitem[{\citenamefont{{Bonanno} et~al.}(2007)\citenamefont{{Bonanno},
  {Drago}, and {Lavagno}}}]{lonanno07}
\bibinfo{author}{\bibfnamefont{L.}~\bibnamefont{{Bonanno}}},
  \bibinfo{author}{\bibfnamefont{A.}~\bibnamefont{{Drago}}}, \bibnamefont{and}
  \bibinfo{author}{\bibfnamefont{A.}~\bibnamefont{{Lavagno}}},
  \bibinfo{journal}{Physical Review Letters} \textbf{\bibinfo{volume}{99}},
  \bibinfo{pages}{242301} (\bibinfo{year}{2007}), \eprint{0704.3707}.

\bibitem[{\citenamefont{{Steinheimer} et~al.}(2009)\citenamefont{{Steinheimer},
  {Mitrovski}, {Schuster}, {Petersen}, {Bleicher}, and
  {St{\"o}cker}}}]{steinheimer09}
\bibinfo{author}{\bibfnamefont{J.}~\bibnamefont{{Steinheimer}}},
  \bibinfo{author}{\bibfnamefont{M.}~\bibnamefont{{Mitrovski}}},
  \bibinfo{author}{\bibfnamefont{T.}~\bibnamefont{{Schuster}}},
  \bibinfo{author}{\bibfnamefont{H.}~\bibnamefont{{Petersen}}},
  \bibinfo{author}{\bibfnamefont{M.}~\bibnamefont{{Bleicher}}},
  \bibnamefont{and}
  \bibinfo{author}{\bibfnamefont{H.}~\bibnamefont{{St{\"o}cker}}},
  \bibinfo{journal}{Physics Letters B} \textbf{\bibinfo{volume}{676}},
  \bibinfo{pages}{126} (\bibinfo{year}{2009}), \eprint{arXiv:0811.4077}.

\end{thebibliography}

\end{document}